\documentclass{emulateapj}
\usepackage{graphicx}
\usepackage{epstopdf}
\usepackage{natbib}
\usepackage{amsmath}
\usepackage{amssymb}

\usepackage{csz} 
\usepackage{abrev} 
\usepackage{sidecap}

\shortauthors{Christensen et al.}

\begin{document}

\title{In-N-Out: the gas cycle from dwarfs to spiral galaxies}

\author{Charlotte R. Christensen}
\affil{Physics Department, Grinnell College, 1116 Eighth Ave., Grinnell, IA 50112, United States}
\email{christenc@grinnell.edu}

\author{Romeel Dav\'e}
\affil{University of the Western Cape, Bellville, Cape Town 7535, South Africa; \\
South African Astronomical Observatories, Observatory, Cape Town 7925, South Africa; \\
African Institute for Mathematical Sciences, Muizenberg, Cape Town 7945, South Africa}

\author{Fabio Governato}
\affil{Astronomy Department, University of Washington, 3910 15th Ave NE, Seattle, WA 98195-0002}

\author{Andrew Pontzen}
\affil{ Department of Physics and Astronomy, University College London,  Gower Street,  London, WC1E 6BT, United Kingdom}

\author{Alyson Brooks}
\affil{Department of Physics and Astronomy, Rutgers University, the State University of New Jersey, 136 Frelinghuysen Road, Piscataway, NJ 08854-8019, United States}

\author{Ferah Munshi}
\affil{Department of Physics and Astronomy, University of Oklahoma, 440 W. Brooks St. Norman, OK 73019, United States}

\author{Thomas Quinn}
\affil{Astronomy Department, University of Washington, 3910 15th Ave NE, Seattle, WA 98195-0002}

\author{James Wadsley}
\affil{Department of Physics and Astronomy, McMaster University, 1280 Main St. W, Hamilton, Ontario, Canada}

\def\path{/Users/christensen/Dropbox/outflows/}
\def\path{./}

\begin{abstract}
We examine the scalings of galactic outflows with halo mass across a suite of twenty high-resolution cosmological zoom galaxy simulations covering halo masses from $10^{9.5}-10^{12}M_\odot$.
These simulations self-consistently generate outflows from the available supernova energy in a manner that successfully reproduces key galaxy observables including the stellar mass-halo mass, Tully-Fisher, and mass-metallicity relations.
We quantify the importance of ejective feedback to setting the stellar mass relative to the efficiency of gas accretion and star formation.
Ejective feedback is increasingly important as galaxy mass decreases; we find an effective mass loading factor that scales as $v_{\rm circ}^{-2.2}$, with an amplitude and shape that is invariant with redshift.
These scalings are consistent with analytic models for energy-driven wind, based solely on the halo potential.
Recycling is common: about half the outflow mass across all galaxy masses is later re-accreted.
The recycling timescale is typically $\sim 1$~Gyr, virtually independent of halo mass.
Recycled material is reaccreted farther out in the disk and with typically $\sim 2-3$ times more angular momentum.
These results elucidate and quantify how the baryon cycle plausibly regulates star formation and alters the angular momentum distribution of disk material across the halo mass range where most of cosmic star formation occurs.

\end{abstract}

\section{Introduction}\label{sec:intro}
\subsection{Evidence for outflows}

Galactic outflows driven by stellar feedback have emerged over the past decade as an integral aspect of how galaxies form and evolve.
Outflows are commonly detected in star-forming galaxies at $z>\sim 0.5$~\citep[e.g.][]{pettini03,Shapley2003,Weiner09,Steidel10,genzel11,rubin13},
and preliminary estimates of the mass loss rate indicate that it is of the same order as the star formation rate~\citep[e.g.][]{erb08,Steidel10,genzel13}.  
The ubiquity of outflows during this cosmic period and their lack of association with active galactic nuclei \citep[e.g.][]{Shapley2003} suggests that they are not associated with
a particular phase of galaxy formation.
Instead, it appears that these galactic outflows are common in galaxies lying on the star forming galaxy main sequence and are likely driven by their high star formation rates or some related property.

Theoretical models of galaxy formation have also found it increasingly necessary to invoke stellar feedback-driven galactic outflows.
Strong and ubiquitous outflows first appeared in galaxy formation simulations in order to explain observations of widespread metal enrichment in the intergalactic medium (IGM) at early epochs~\citep{aguirre01,Oppenheimer2006}.  
However, it was soon recognized that these same outflows had a considerable impact on the properties of galaxies themselves, such as their stellar and baryonic content, their metallicity, their mass distribution, and the state of the circum-galactic medium (CGM).
Today, the majority of successful galaxy formation models, be they hydrodynamic or semi-analytic~\citep[e.g.][]{Somerville2015}, include stellar feedback-driven galactic outflows as a central feature.

\subsection{Impact of outflows on galaxies}
Galactic outflows have a strong impact on the growth of the stellar and gaseous content of galaxies.  
First, mass loss limits the amount of baryons in the disk of galaxies, thereby impacting the observed baryonic mass fraction within the halos of galaxies \citep[e.g.][]{dave09} and the total stellar and disk gas mass of the galaxies \citep[e.g.][]{scannapieco11}.  
Mass loss, in combination with heating of the interstellar media by feedback \citep[e.g.][]{brooks07}, also impacts the observed galaxy stellar mass function by limiting the amount of star formation in the disk of the galaxy \citep{SpringelANDHernquist03b,Dave11}.  
By reducing the amount of gas available for star formation, outflows mitigate the overcooling problem wherein too large a fraction of baryons condenses into stars \citep{WhiteANDReese78, Balogh01, Dave01}.  
Moreover, stellar feedback-driven outflows are frequently invoked to explain the sub-L* portion of the stellar mass-halo relation \citep[e.g.][]{Shankar2006} and the simulations most successful in reproducing the stellar mass-halo mass relation typically employ strong stellar feedback \citep{Stinson13,Aumer2013,Hopkins2013}.
The efficiency of outflows across a range of galaxy masses is therefore key to determining the stellar and baryonic mass for different mass halos.

As well as globally impacting the baryons within galaxies, outflows play an important role in determining the distribution of matter.  
For example, the stellar feedback that drives outflows delays star formation, which results in less angular momentum loss during galaxy mergers and, therefore, less centrally concentrated galaxies \citep{WhiteANDReese78, Dekel86,  navarrosteinmetz97, robertson04, Okamoto05, Governato09, Scannapieco08}.  

In addition to being an agent for reducing angular momentum loss, stellar feedback is now also considered a mechanism for removing low-angular momentum baryons in galaxies.
This removal is necessary for the modeling of realistic galaxies: while the overall specific angular momentum of disks is comparable to that of their parent dark matter halos \citep{fall80,dalcanton97,mo98}, the observed {\em distribution} of the specific angular momentum in any given disk differs from predictions based on pure angular momentum conservation.  
In particular, observed galaxies are deficient in low-angular momentum material compared to simple disk collapse models~\citep{VanDenBosch01a}.
Strong galactic winds were first posited as a mechanism for removing low-angular momentum material in \citet{Binney01} and recent simulations have demonstrated that stellar feedback preferentially removes low-angular momentum material from galaxies \citep{Governato10, Brook11a, Brook12a, Maccio12, Ubler2014}.  
The loss of low-angular momentum baryons results in bulgeless dwarf galaxies \citep{Governato10}, and spiral galaxies with more realistic central baryonic distributions \citep{Brook11b, Christensen14, AnglesAlcazar13}.

While outflows are typically thought of as a way to remove material, the reaccretion of that material (wind recycling) also impacts galaxy evolution.
Some outflowing material must be transferred to the diffuse IGM in order to explain observations \citep[e.g.][]{Cowie1995} of metal-line absorption \citep{Oppenheimer2006,Cen2011,Oppenheimer2012}. 
However, in many cases the outflows are thought not to escape the CGM, but to instead return to the galaxy on relatively short timescales.
This so-called wind recycling adds to the pristine accretion from the IGM and to the accretion of gas already bound in galaxies (i.e. mergers).  
The recycling of previously ejected wind material can be a key factor in setting the galaxy stellar mass function~\citep[e.g.][]{Oppenheimer10,Bower2012}.  
It is also thought that fountaining gas gains angular momentum through interactions within the halo environment before being reaccreted \citep{Marasco2012}, which further shifts the angular momentum distribution of the disk baryons to higher values \citep{Brook11b}.

\subsection{The analysis of outflows presented here}
Together, inflows, outflows, and wind recycling govern many of the key physical properties of galaxies~\citep[e.g.][]{Dave11b}.
Therefore, it is critical to understand the operation of the baryon cycle, including the scaling of the mass loading factor with galactic properties, the relative rates of outflowing, accreted, and recycled gas, and the source and eventual destination of outflowing material.  
Hydrodynamic simulations of galaxy growth are a valuable tool for this, because the inherently dynamical nature of the baryon cycle requires a fully dynamical model to capture it properly.  
Such simulations must be set within a fully cosmological context, since the accretion is cosmologically-driven. 
Furthermore the details of the complex interactions between inflows, stellar feedback, outflows, and ambient halo gas strongly motivates very high numerical resolution.  
These requirements are a challenge for current galaxy formation models, one that is only recently starting to be met using cosmological ``zoom" simulations, in which an individual galaxy is re-simulated at much higher resolution within a larger cosmological volume.

In this paper we investigate the detailed dynamics of inflow and outflow processes using a suite of cosmological zoom simulations.
We take advantage of the particle-based nature of our Smoothed Particle Hydrodynamics (SPH) simulations run with the {\sc Gasoline} code to directly track all mass movement in and out of the disk and CGM.  
This particle tracking enables us to directly study recycling and to identify the source and future trajectory of individual parcels of gas.
We analyze high-resolution simulations of twenty spiral and dwarf galaxies that span two and a half orders of magnitudes in virial mass, all simulated with the same physics and comparable numerical resolution.  
Our work improves on previous particle tracking analyses that have focused either on low-resolution (non-zoom) simulations of many galaxies \citep{Oppenheimer10} or on a few simulations of similar-mass galaxies \citep{Brook11b,Woods2014,Ubler2014} and complements non-particle tracking studies of galactic winds \citep{Muratov2015}.

Our simulations self-consistently generate outflows from the available supernova energy using a methodology that has been shown to successfully reproduce a wide variety of galaxy observations.
In this method, the transfer of stellar feedback energy depends only on the local properties of the ISM, and since the feedback model is ignorant of the host galaxy properties, outflow trends with mass result from the dynamics of the simulation.  
This analysis enables us to independently study inflows, outflows, and recycling as a function of galaxy mass, which allows a deeper investigation into the underlying physical processes that govern the baryon cycle.
The details of baryon cycling presented here do depend on our methodology for driving outflows, and may vary for different choices of physical models~\citep[see e.g.][]{Keller2015,Muratov2015}.  Nonetheless, our results are of particular interest since our outflow driving model yields a viable match to numerous key observational constraints, as we will show.

Our paper is organized as follows.
In \S\ref{sec:methods} we describe the simulations used and the particle tracking analysis used to determine outflows.
We justify our simulation models in \S\ref{sec:globalprop} by comparing their global properties to observed trends.
In \S\ref{sec:baryfrac} -- \ref{sec:massloading}, we analyze the efficiency of various forms of feedback across galaxy mass.
We examine gas recycling and the characteristics of reaccreted material \S\ref{sec:vel}  -- \ref{sec:wherewhat}.
We discuss our results in light of other models and numerical concerns in \S\ref{sec:discuss}.

\section{Simulation and Analysis}\label{sec:methods}

We compared the properties of outflows across a set of twenty field
galaxies with final virial masses between $10^{9.5}$ to $10^{12}\Msun$
that were simulated using the $N$-Body + SPH  code, {\sc gasoline} \citep{Wadsley04}.
{\sc gasoline} is an SPH extension to {\sc pkdgrav}
\citep{stadel01}, a parallel, gravity-tree based $N$-Body Code.
The simulations were integrated to a redshift of zero in a fully-cosmological, $\Lambda$CDM context using WMAP3 \citep{Spergel07} parameters: $\Omega_0$=0.24, $\Lambda$=0.76, h=0.73, $\sigma_8$=0.77. 

All twenty galaxies are central galaxies selected from a set of seven simulations.
In order to achieve significantly higher resolution while still
modeling the effects of the large-scale environment, we used the
``zoom-in'' volume renormalization technique \citep{katz93}.
More specifically, to create the initial conditions for these
simulations we selected seven field-like regions from two uniform
dark matter-only simulations, one representing a 25$^3$ Mpc$^3$ and
the other a 50$^3$ Mpc$^3$ volume.  These regions (~0.05\% of the total volume) were then resimulated
at higher resolution in the context of the larger dark matter-only simulation.  
The final sample galaxies had massive dark particle contamination of less than 0.07\% their total dark matter mass and 11 of the galaxies were completely free from contamination.
The force spline softening lengths are either $\epsilon=87$ or 170~pc, and the particle
masses for the dark matter, gas, and stars (at their formation)
are, respectively, 1.6 (13)$\times 10^4$, 3.3 (27.0)$\times$10$^3$,
and 1.0 (8.0)$\times$10$^3$M$_{\odot}$.  The simulations have a minimum
smoothing length of $0.1\epsilon$, which is sufficient to resolve
the disks of galaxies and the giant molecular clouds within which stars
form. 
All the simulations used a force accuracy criterion of $\theta$ = 0.725 and a Courant condition of $\eta_C$=0.4.
Particle time steps were required to satisfy $\Delta t = \eta \sqrt{ \frac{\eta_i}{a_t}}$ where $\eta_i$ is the particle's gravitational softening, $a_i$ is the particle's acceleration, and $\eta$=0.195.
The parameters used to simulate the galaxies and their final properties at $z=0$ are listed in Table 1.

\begin{deluxetable*}{lcccccccc}   
\tablecaption{
Properties of the galaxies at z = 0.
}
\tablecolumns{9} 
\tablewidth{0pt} 
\tablehead{
\colhead{Simulation} &
\colhead{Softening} &
\colhead{Gas Particle} &
\colhead{Halo ID} &
\colhead{Virial Mass} &
\colhead{Gas Mass} &
\colhead{Cold Gas} &
\colhead{Stellar } &
\colhead{$V_f$}\\
\colhead{Name}  &
\colhead{Length} &
\colhead{Mass}	 &
\colhead{		}   &					
\colhead{		} &
\colhead{in $R_{vir}$ } & 
\colhead{Mass} &
\colhead{Mass} \\ 
\colhead{ } &
\colhead{[pc]} &	
\colhead{[$\Msun$]} &	
\colhead{ } &
\colhead{[$\Msun$]}	 &
\colhead{[$\Msun$]}	 &
\colhead{[$\Msun$]} 	 &
\colhead{[$\Msun$]} 	  &
\colhead{km/s]}}
\startdata
				& (1)			& (2)				& (3)				& (4)					& (5)						& (6)					& (7)					& (8) 		\\\hline \hline
h799 			& 87			& $3.3 \times 10^3$	& 1$^{2,3,5}$		& $2.4\times 10^{10}$	& $1.4\times 10^{9}$			& $2.5\times 10^{8}$		& $1.4\times 10^{8}$		& 55			\\
				&			&				& 4				& $6.8\times 10^{9}$		& $4.1\times 10^{7}$			& $1.0\times 10^{7}$		& $1.8\times 10^{7}$		& 33			\\ 
				&			&				& 6				& $4.4\times 10^{9}$		& $3.9\times 10^{7}$			& $2.6\times 10^{7}$		& $3.5\times 10^{6}$		& 27			\\ 
\hline
h516	 			& 87			& $3.3 \times 10^3$	& 1$^{1,2,3,5}$		& $3.8\times 10^{10}	$	& $2.3\times 10^{9}$			& $5.5\times 10^{8}$		& $2.5\times 10^{8}$		& 67			\\
			 	&			&				& 2		 	   	& $1.5\times 10^{10}	$	& $3.7\times 10^{8}$			& $4.6\times 10^{7}$		& $8.1\times 10^{7}$		& 34			\\
\hline
h986			 	& 170		&  $2.7 \times 10^4$	& 1$^{3,5}$  		& $1.9\times 10^{11}$	& $1.7\times 10^{10}$		& $3.5\times 10^{9}$		& $4.5\times 10^{9}$		& 103		\\
			 	&			&				& 2		 	   	& $5.9\times 10^{10}	$	& $3.2\times 10^{9}$  		& $7.4\times 10^{8}$		& $1.2\times 10^{9}$	  	& 77			\\
				&			&				& 3		 	   	& $3.8\times 10^{10}	$	& $2.4\times 10^{9}$ 		& $5.4\times 10^{8}$		& $4.6\times 10^{8}$	  	& 76			\\
				&			&				& 8		 	   	& $1.1\times 10^{10}	$	& $6.4\times 10^{7}$ 		& $1.3\times 10^{7}$  	& $4.0\times 10^{7}$	  	& 35			\\
				&			&				& 15				& $4.4\times 10^{9}	$	& $8.7\times 10^{7}$ 		& $2.7\times 10^{8}$ 	& $6.2\times 10^{6}$		& 29			\\
				&			&				& 16				& $3.2\times 10^{9}	$	& $3.0\times 10^{7}$ 		& $1.1\times 10^{8}$		& $2.3\times 10^{6}$		& 27			\\
\hline
h603			 	& 170		& $2.7 \times 10^4$	& $1^{3,5}$ 		& $3.4\times 10^{11}$	& $3.1\times 10^{10}	$		& $4.2\times 10^{9}$		& $7.8\times 10^{9}$		& 115		\\ 
			 	&			&				& $2^3$		 	& $1.0\times 10^{11}	$	& $6.1\times 10^{9}$  		& $7.8\times 10^{8}$  	& $3.8\times 10^{9}$	  	& 75			\\
				& 			&				& 3		 	   	& $2.9\times 10^{10}	$	& $1.8\times 10^{8}$			& $1.8\times 10^{8}$  	& $3.9\times 10^{8}$  	& 50			\\
\hline
h258			 	& 170		& $2.7 \times 10^4$	& $1^{3,4}$		& $7.7\times 10^{11}	$	& $5.6\times 10^{10}$ 		& $5.7\times 10^{9}$  	& $4.5\times 10^{10}	$  	& 182		\\
			 	&			&				& 4		 	   	& $1.1\times 10^{10}	$	& $1.4\times 10^{8}$  		& $6.2\times 10^{7}$  	& $5.9\times 10^{7}	$  	& 43			\\
\hline
h285			 	& 170		& $2.7 \times 10^4$	& $1^{3}$			& $8.8\times 10^{11}	$	& $6.3\times 10^{10}$  		& $8.5\times 10^{9}$  	& $4.6\times 10^{10}	$  	& 164		\\
			 	&			&				& 4		 	   	& $3.4\times 10^{10}	$	& $1.2\times 10^{9}$  		& $1.5\times 10^{8}$  	& $3.9\times 10^{8}$  	& 64			\\
				&			&				& 9		 	   	& $1.2\times 10^{10}	$	& $3.1\times 10^{8}$  		& $1.3\times 10^{8}$  	& $5.4\times 10^{7}$  	& 52			\\
\hline
h239			 	& 170		& $2.7 \times 10^4$	& $1^{3}$		 	  & $9.1\times 10^{11}	$	& $8.1\times 10^{10}$  		& $6.2\times 10^{9}$  	& $4.5\times 10^{10}$  	& 165	
\enddata
\tablecomments{ Column 5 lists the total mass of gas within the virial radius while column 6 lists the mass of only HI, $\Hmol$, and HeI. The stellar mass listed in column 7 is calculated directly from the simulation.}
\tablenotetext{1}{Appears in \citet{Christensen12}.}
\tablenotetext{2}{Appears in \citet{Governato12}.}
\tablenotetext{3}{Appears in \citet{Munshi12}.}
\tablenotetext{4}{Appears in \citet{Zolotov12}.}
\tablenotetext{5}{Appears in \citet{Christensen12a}.}

\end{deluxetable*}


We integrate over the H and He chemical networks to produce
non-equilibrium ion abundances and $\Hmol$ abundance \citep{Christensen12}.
$\Hmol$ forms both on dust grains, assuming a fixed dust-to-metallicity
ratio and a clumping factor of 10~\citep{Wolfire08, Gnedin09} and
via H$^-$, following the minimal model of \citet{Abel97}.
Photo-ionization and heating rates of H and He are calculated assuming a set redshift-dependent cosmic ultraviolet (UV) background, (Haardt \& Madau 2005)\footnote{ Haardt \& Madau (2005) refers to an unpublished updated version of \citet{Haardt96}, specified in {\sc cloudy} \citep{Ferland98} as ``table HM05.''}.
The Lyman-Werner radiation, which is responsible
for $\Hmol$ photodissociation, is calculated based on emission from
nearby stellar particles \citep{Christensen12}.  $\Hmol$ is shielded
from dissociating radiation both through self-shielding and dust
shielding \citep{Draine96, Glover07, Gnedin09}, using the smoothing
lengths of particles for the column lengths.  Similarly, HI is
shielded from photo-ionizing radiation by dust.

Cooling channels include collisional ionization \citep{Abel97},
$\Hmol$ collisions, radiative recombination \citep{Black81,
VernerANDFerland96}, photoionization, bremsstrahlung, and HI, $\Hmol$
and He line cooling \citep{Cen92}.  Additional cooling takes place
via metal lines \citep{Shen10}.  The metal line cooling rates
used in the code are tabulated based on the gas temperature,
density, and metallicity and the cosmic UV background.  These rates
are computed using {\sc cloudy} \citep[version 07.02;][]{Ferland98}
under the assumptions that the gas was in ionization equilibrium
and optically thin to UV radiation.

{\sc gasoline} separately follows both the oxygen and iron abundances of gas
particles and the total metal production.  These metals are injected into the gas by type I and
II SN following \citet{Raiteri96} and distributed across the smoothing
sphere.  They are also injected by stellar
winds using a model for mass loss that follows \citet{Weidemann87}
and assumes that the metallicity is that of the stellar particle.
Metals are further distributed throughout the gas by diffusion
\citep{Shen10}.


Star formation proceeds probabilistically, according to the gas
density and $\Hmol$ fraction.  The dependency on $\Hmol$ represents
the observed connection between the star formation rate and the
local $\Hmol$ abundance \citep[e.g.][]{Bigiel08}.  The probability
$p$ of a given gas particle forming a star is
\begin{equation}
 p = \frac{m_{gas}}{m_{star}}(1 - e^{-c^* \frac{X_\Hmol}{X_{\Hmol} + X_{\mathrm{HI}}} \Delta t /t_{dyn}})
\end{equation}
where $m_{gas}$ is the mass of the gas particle, $m_{star}$ is
the mass of the potential star, $c^*=0.1$ is the star forming efficiency,
$X_\Hmol$ and $X_\mathrm{HI}$ are the mass
fractions of the particle in the form of $\Hmol$ and HI, respectively,
$\Delta t$ is the time step, and $t_{dyn}$ is the dynamical time.
Star formation is allowed to proceed only in gas particles that
are denser than 0.1 amu cm$^{-3}$ and colder than 10$^3$ K.  However,
the dependency on the $\Hmol$ abundance made these two constraints
largely irrelevant since almost all stars form at gas densities greater than 10 amu cm$^{-3}$.


Supernova feedback is incorporated using the ``blastwave'' approach
\citep{Stinson06}, in which the theoretical solution to a blastwave
explosion in a medium of a given density and pressure are used to
determine the spatial extent of the feedback.  In this
approach, energy is distributed to nearby gas particles that lie
within the maximum radius of the SN blastwave \citep{Chevalier74}.
The cooling of these affected particles is disabled for a period
of time equal to the theoretical lifetime of the hot, low-density
shell produced during the momentum-conserving phase of the supernova
remnant \citep{McKee77}.  Typically, these periods of time are on
the order of several times $10^7$ years.  The total amount of energy
deposited in the ISM is the canonical 10$^{51}$ ergs per supernova.

The blastwave recipe differs from many other subgrid feedback recipes \citep[e.g.][]{SpringelANDHernquist03a,Dave11,scannapieco11} in that wind particles are not provided with an artificial kick.
Even when cooling is disabled for the feedback-affected gas particles, they continue in all other ways to interact hydrodynamically with the rest of the simulation.
Additionally, the feedback depends only on the local gas properties;
it is independent from the large-scale properties of the galactic
halo.  This feedback recipe does not include a separate model for
other forms of stellar feedback, such as radiation pressure, which
may help drive a galaxy wind either by adding additional momentum
to the gas, or causing the gas to be
more responsive to supernova feedback~\citep{Hopkins2013,Stinson13}.
As such, this efficient transfer of supernova energy into the ISM
and the temporary delay of cooling is best interpreted as a model
for the {\em total stellar} feedback from all processes related to
young stars.  In \S~\ref{sec:discuss}, we discuss the possible
impacts of excluding other forms of stellar feedback.

\subsection{Post-processing Analysis}

Individual halos are identified in each snapshot using {\sc amiga's halo finder} \citep{Knollmann2009,Gill2004} \footnote[1]{{\sc amiga's halo finder} is available for download at http://popia.ft.uam.es/AHF/Download.html}, which uses a grid hierarchy to identify areas of over-density and iteratively removes gravitationally unbound particles from the prospective halos.
The virial radius, $R_{vir}$, is defined to be the radius for which the average halo density is some multiple of the background density.
This value for the average halo density evolves with redshift but is approximately equal to 100 times the critical density. 
In determining the evolution of the halo, we use a merger tree to trace the main
progenitor back in time.  At each snapshot, the main progenitor is
defined to be the halo in the previous step that contains the
majority of the particles in the current halo.

\begin{figure*}
\begin{center}
\includegraphics[width=1\textwidth]{\path 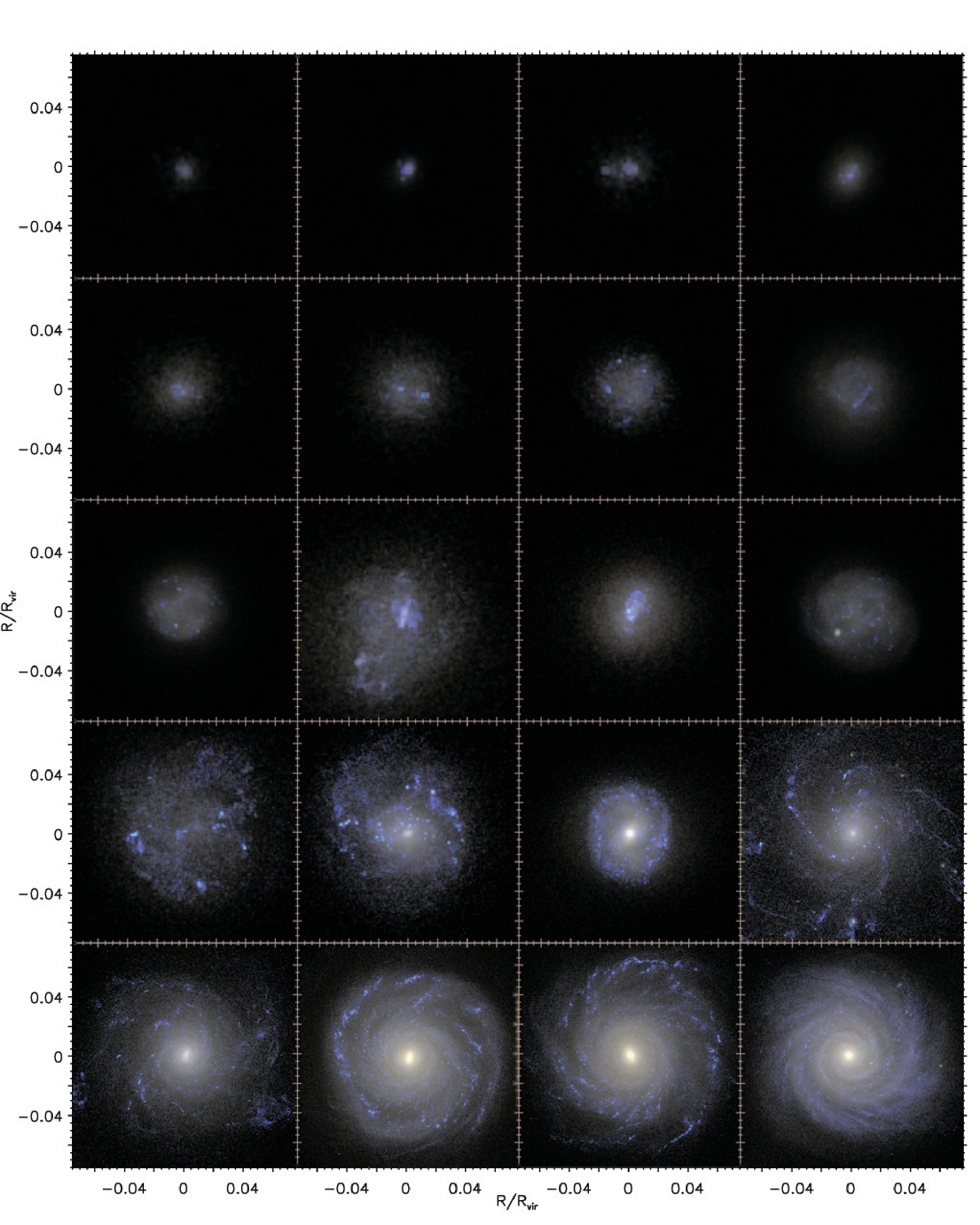} 
\end{center}
\caption[Mock Observations]
{ 
Simulated observations of the sample of galaxies in SDSS $r$, $g$, and $i$ bands at z = 0 and ranked by mass.
All galaxies are shown face-on and the images were generated using {\sc sunrise}.
}
\label{fig:sunrise}
\end{figure*}

In order to verify the properties of the simulated galaxies, we
make as direct a comparison to observations as possible (e.g. see
\S~\ref{sec:globalprop} for the comparisons to the stellar mass-halo mass, Tully-Fisher and
mass-metallicity relationships).  This comparison requires the
generation of mock-photometric magnitudes, which was accomplished
using the ray tracing radiative transfer program, {\sc sunrise}
\citep{Jonsson06}.  
Magnitudes in different bands were calculated for the galaxy oriented at a 45$^\circ$ angle with the midplane defined by the angular momentum axes of the gas within 5 kpc of the center (comoving).
Face-on {\sc sunrise}-generated images of all the
galaxies in SDSS $g$, $r$, and $i$ filters are shown in
Figure~\ref{fig:sunrise}.

Table 1 lists the global properties of the halos, including virial,
total gas, cold gas, and stellar mass for each of the halos at a
redshift of zero.  The stellar mass listed is {\bf the total stellar mass within the virial radius} and is calculated directly from the simulation (rather than from mock-photometric observations). We list both the total gas mass (mass of gas
particles within $R_{vir}$) and cold gas mass in order to distinguish
between the entirety of the gas mass and the fraction of it that
is easily observable.  Here the cold gas mass was defined to be
gas mass in the form of HI, $\Hmol$, and HeI.

\subsubsection{Particle Tracking}

In order to study the baryon cycle within galaxies in detail, the
gas must be followed as it is accreted to and is ejected from the disk.  As
such, we determine gas accretion and gas loss through particle
tracking. Essentially, we divide the gas between disk, halo, and IGM for
each of the snapshots, and then the movement between these phases marks
inflow and outflow.  In addition to allowing us to identify gas
recycling, this particle tracking enables us to determine the source
of gas outflows.  

We used particle tracking to identify {\em accreting} and {\em outflowing} gas and further subdivided the outflowing gas into {\em ejected} gas that became dynamically unbound from the disk and {\em expelled} gas that escaped the halo.
Our method for identifying these incidents is as follows.
First, at each snapshot we determine the gas particles that are in
the main halo and the disk of the galaxy.  Gas particles are
considered part of the galaxy if the halo finder determined them
to be a member of either the main halo or one of its satellites.
Gas particles are considered part of the disk of the galaxy during a snapshot if they
meet all of the following criteria:
\begin{itemize}
\item has a density $n\geq0.1$~amu~cm$^{-3}$
\item has a temperature $T\leq 1.2 \times 10^4$~K
\item are within 3 kpc of the midplane of the galaxy.
\end{itemize}
The density and temperature cuts select for the cool interstellar medium 
phase of gas, while the spatial cut excludes the ISM of satellite galaxies.
For this analysis, we focus on tracking particles since $z=3$. 

Instances of gas {\it accretion} includes both the first snapshot a gas particle is identified as part of the disk, and each time it reenters the disk.

{\em Ejected} gas is defined to be the gas particles which not only stop being identified as part of the disk but also become gravitationally unbound from the $baryonic$ disk of the galaxy.
These ejected particles have, at some snapshot subsequent to leaving the disk, a kinetic energy greater than the gravitational potential from the combined disk gas plus stellar mass (calculated as if all the mass were located at the center of mass of the galaxy).
Note that these particles need not be unbound from the greater dark matter potential well.
Gas particles may be ejected multiple times in their history provided they are identified as part of the disk in between ejection events.

While ejected gas particles need only become unbound from the baryonic disk, a subset of the ejected gas eventually also escapes the halo.
Any gas particle that ends up beyond $R_{\rm vir}$ at a snapshot subsequent to ejection is identified as having been expelled from the halo.

In addition to the ejected gas, there is a larger population of gas particles that are labeled part of the disk in one snapshot, but not in a subsequent one.  
This occurs whenever a gas particle becomes too hot, too low density or too distant from the mid-plane of the galaxy.
As such, it can include tidally stripped material, in addition to gas affected by feedback.
Ejected gas particles, therefore, are a subsection of this broader category just as the gas particles that are expelled from the halo are a subsection of the ejected gas.

These identifications of gas outflow and accretion are limited by the temporal spacing of the snapshots.  
For the simulations in this paper, snapshots were spaced approximately 100 Myrs apart.  
Our time resolution for tracing inflows and outflows, therefore, is also about 100 Myr.  
Furthermore, the limited number of snapshots mean that particles may have been ejected and reaccreted between two concurrent snapshots.  
In such a case, an outflow would not have been identified.  
This limited time resolution means that the outflowing and accretion rates must be considered lower limits.
The similar snapshot spacing across simulations ensures, however, that we can draw a comparisons across the different galaxies.

\begin{figure*}
\begin{center}
\includegraphics[width=1\textwidth]{\path 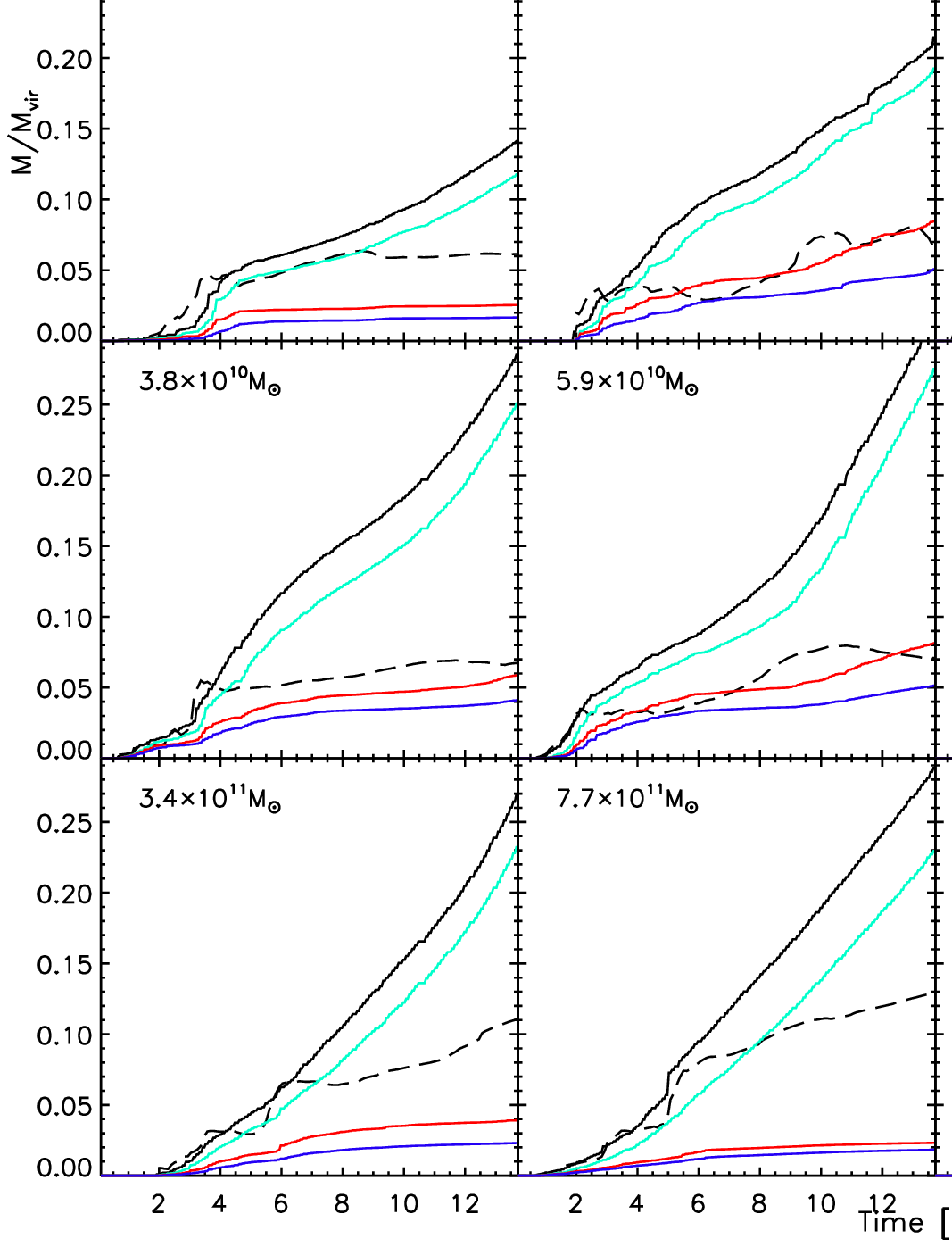} 
\end{center}
\caption[Outflow history]
{ 
The cumulative mass loss history for each of the galaxies.
In these figures, the mass loss is scaled by the final virial mass of the galaxies.
The solid black line indicates the total cumulative mass of gas particles accreted onto the disk of the galaxy, including any reaccretions of the same particle.
The colored solid lines show the cumulative mass of gas particles ever removed from the disk (green), ejected such that they become dynamically unbound from the disk (red), and expelled beyond the virial radius (blue).
For comparison, the dashed line show the total baryonic mass within the disk.
Occasionally the dashed line (total disk baryon mass) lies above the black solid line (cumulative mass of accreted gas) because of stellar accretion.
}
\label{fig:outflow_history}
\end{figure*}

Figure~\ref{fig:outflow_history} shows the cumulative mass loss
history for each of our simulated galaxies versus time since the
Big Bang, as a fraction of the final virial mass (listed in the
upper left).  The dashed line shows the total baryonic mass in the
disk as a fraction of the redshift zero halo mass.  Merger
events can be identified by the sudden jumps, more common in larger-mass
halos and at earlier epochs.

The black line shows the cumulative gas mass accreted to the disk, including reaccretion events (stars are not included).
The green line shows the cumulative outflow mass from the disk.  While
disk masses tend to stabilize at later epochs, the total mass loss generally increases roughly linearly with time, mimicking the accreted mass.  
According to these accretion and outflow definitions, a roughly constant mass of gas enters and leaves the disks at late times.
Also notable is the greater amounts of gas accretion and loss from the disk compared to the final virial mass in more massive galaxies.
The smaller baryonic fraction of dwarf galaxies is the result of both lower accretion rates compared to their virial mass and the loss of a greater fraction of their accreted material, shown further in \S~\ref{sec:baryfrac}.

We further subdivide the cumulative mass loss into gas ejected from the disk (red) and gas that is expelled from the halo (blue).  
Since gas expelled from the halo is a subset of gas ejected from the disk, it is always lower, but the two generally track each other well.  
At late times much of the disk mass loss is not in ``ejected'' material (i.e., gas that becomes dynamically unbound from the disk).
Rather, it is dominated by gas that escapes the disk but does not become dynamically unbound.  
This gas that is removed from the disk but not ejected includes material heated by supernovae to above $10^4$K, gas that is tidally stripped, and gas that is entrained by outflowing gas.

\section{Results}\label{sec:results}
\subsection{Global Galaxy Properties}\label{sec:globalprop}

The balance between gas accretion, star formation, and outflows determines the baryonic content of galaxies.  Therefore, the observed stellar and cold gas masses of galaxies act as a basic constraint on theoretical models.
Our first step, therefore, is to verify our simulated galaxies' agreement with global $z=0$ trends in the baryonic content of observed galaxies, in particular the stellar mass-halo mass, Tully-Fisher, and mass-metallicity relations.

\begin{figure}
\begin{center}
\includegraphics[width=0.5\textwidth]{\path 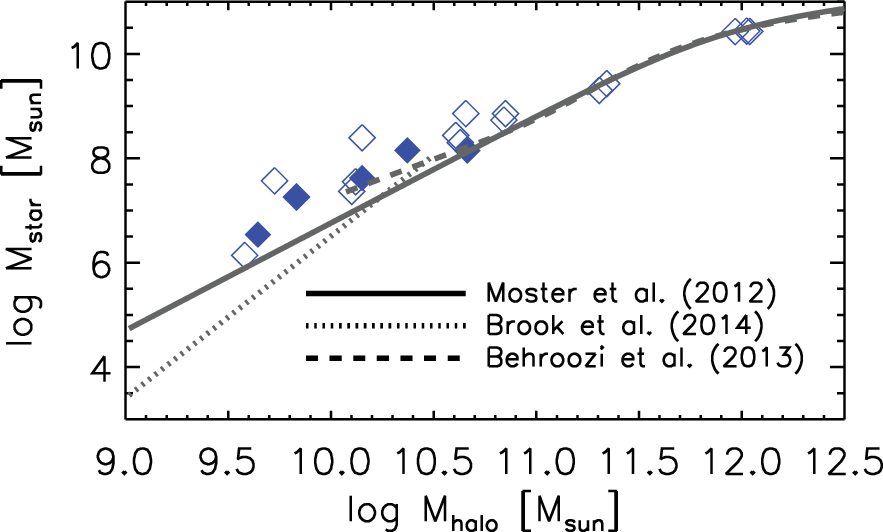}
\end{center}
\caption[Stellar Mass Halo Mass]
{ 
The redshift zero stellar mass-halo mass relation for the simulated galaxies (blue diamonds), compared to abundance matching-derived fits from \citet{
Behroozi2013}, \citet{Moster13} and \citet{Brook2014} (grey lines).
Filled diamonds represent high resolution ($\epsilon=85$~pc) simulations, while empty diamonds represent the medium-resolution ($\epsilon=170$~pc) simulations.
In order to better mimic the observations, stellar masses were determined from mock-photometric observations and halo masses were taken from dark matter-only simulations.
}
\label{fig:smhm}
\end{figure}

The halos in our sample are shown to match the $z=0$ stellar-mass halo-mass relation inferred from abundance matching \citep{Behroozi2013, Moster13, Brook2014} in Figure~\ref{fig:smhm}.
In order to make an accurate comparison, stellar masses were calculated from simulated photometric observations while halo masses were taken from dark matter-only simulations, as in \citet{Munshi12}.
The agreement between the simulations and the abundance matching models indicates that the simulated galaxies are able to form the correct mass of stars for their halo mass over the course of their lifetime.
Note that calculating the stellar masses from photometric observations results in smaller masses than measuring them directly from the simulations \citep{Munshi12}.
For these galaxies the actual stellar masses (listed in Table 1) are about 1.3 -- 1.4 times the photometrically determined ones.

\begin{figure}
\begin{center}
\includegraphics[width=0.4\textwidth]{\path 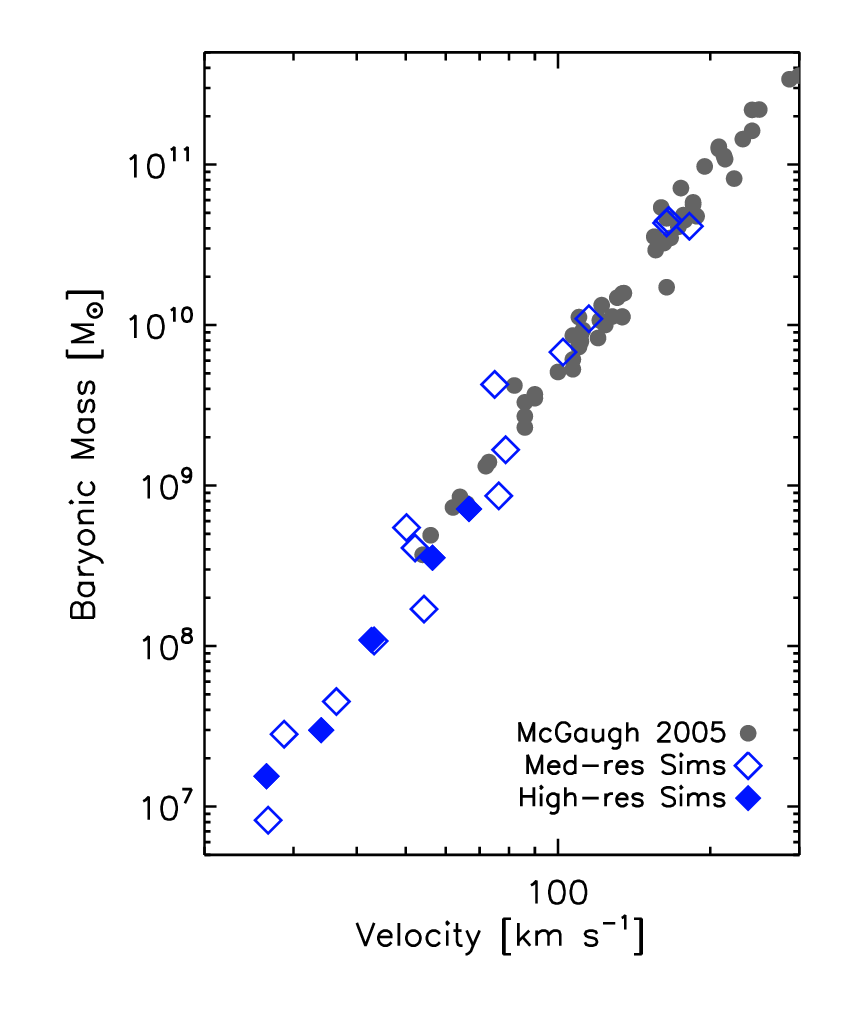}
\end{center}
\caption[Baryonic Tully Fisher Relationship]
{ 
The baryonic Tully Fisher Relation for the simulated galaxies (blue diamonds), compared to observed galaxies from \citet{McGaugh05} (grey circles).
Filled diamonds represent high resolution ($\epsilon=85$~pc) simulations, while empty diamonds represent the medium-resolution ($\epsilon=170$~pc) simulations.
As in the observations, the total baryon mass of the simulated galaxies was calculated from the HI, $\Hmol$ and HeI gas masses and the photometrically-determined stellar mass.
The simulated galaxies lie along the same line as the observed galaxies, indicating that their baryonic masses scale appropriately with their rotational velocities.
}
\label{fig:btf}
\end{figure}

Figure~\ref{fig:btf} compares the simulations to the observed baryonic Tully-Fisher relationship from \citet{McGaugh05} (the sum of the stellar and cold gas mass of galaxies as a function of their rotational velocities).  
To ensure a fair comparison to these data, we compute the baryonic mass as the sum of the stellar mass calculated from the galaxies' B-band magnitudes using a mass-to-light ratio determined by the B-V color following\citet{McGaugh05}, together with the HI, $\Hmol$ and HeI gas mass.
Additionally, the asymptotic velocity, $V_f$, is calculated by fitting either an increasing or decreasing arctangent function to the circular velocities for radii that lay between twice the softening and less than the radius containing 90\% of the cold gas.  
The agreement is excellent over the entire range of overlapping masses.

\begin{figure}
\begin{center}
\includegraphics[width=0.4\textwidth]{\path 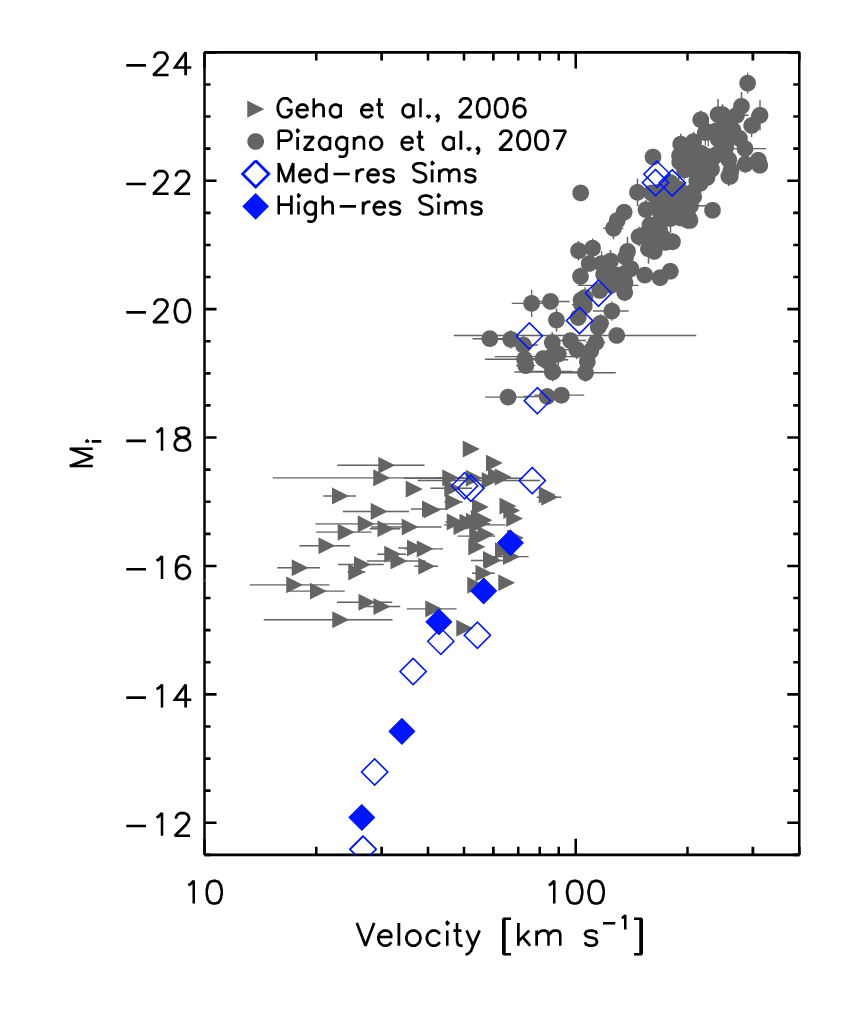}
\end{center}
\caption[Tully Fisher Relationship]
{
The $i$-band Tully-Fisher relation for the simulated galaxies (blue diamonds) compared to observed galaxies 
from \citet{Pizagno2007} and \citet{Geha06} (filled grey circles and squares, respectively).
Filled diamonds show the high resolution simulations, empty diamonds
the medium-resolution runs.
While the \citet{Pizagno2007} data are of the tangental rotational velocities, the data from \citet{Geha06} is of HI line widths.
Therefore, when comparing to the simulated galaxies the data from \citet{Geha06} should be considered lower-limits on the actual tangental rotational velocities.
In general, the simulated galaxies follow the observed Tully-Fisher relation across nearly an order of magnitude in velocity.
}
\label{fig:tf}
\end{figure}

Analogously, Figure~\ref{fig:tf} shows the simulated galaxies along the observed SDSS $i$-band Tully-Fisher relationship from \citet{Geha06} and \citet{Pizagno2007}.  
The \citet{Pizagno2007} data are for the tangental rotational velocities, while the data from \citet{Geha06} employs the HI line width and thus should be interpreted as lower limits to the circular velocities.  
The simulated galaxies follow both the baryonic and standard Tully Fisher relations across nearly an order of magnitude velocity range.

The Tully-Fisher plots confirm that the feedback mechanism in these simulations is able to solve the so-called angular momentum crisis in galaxies noted in early simulations without strong stellar-driven feedback \citep{Navarro00,Steinmetz1999}.  
Feedback primarily solves this problem by the strong suppression of star formation in dwarf galaxies \citep{Munshi12}.
As a result of this suppression, there is less growth via merging that would otherwise generate an overly peaked central rotation curve~\citep{Governato09}.  
Additionally, as will be further discussed in \S\ref{sec:wherewhat}, these outflows reduce the amount of material in the centers of galaxies through the preferential removal of low-angular momentum gas.
As noted by \citet{Brook11b}, outflows play a key role in setting the angular momentum distribution in
galaxies, and the agreement with the baryonic and $i$-band T-F relation suggests that the outflows in these simulations are plausible.

The gas-phase metallicities of galaxies are controlled by the balance between the accretion of pristine gas, the injection of metals into the ISM from stars, the ejection of metal-enriched ISM gas by feedback and the re-accretion of enriched gas \citep[e.g.][]{Finlator08,Peeples11,Dave11b}.  
Despite the complexity of the processes, galaxies follow a well-defined mass-metallicity relationship \citep{tremonti04}, albeit subject to calibration uncertainties in metallicity indicators \citep{Kewley08}.
Thus the mass-metallicity relationship provides another strong test of the plausibility of our outflow model.  
Previously, \citet{brooks07} found galaxies generated using an earlier version of {\sc gasoline} follow the observed mass-metallicity relationship.  
Here, we update that analysis to verify our sample of simulated galaxies with the current ISM model and star formation and feedback parameters.

\begin{figure}
\begin{center}
\includegraphics[width=0.5\textwidth]{\path 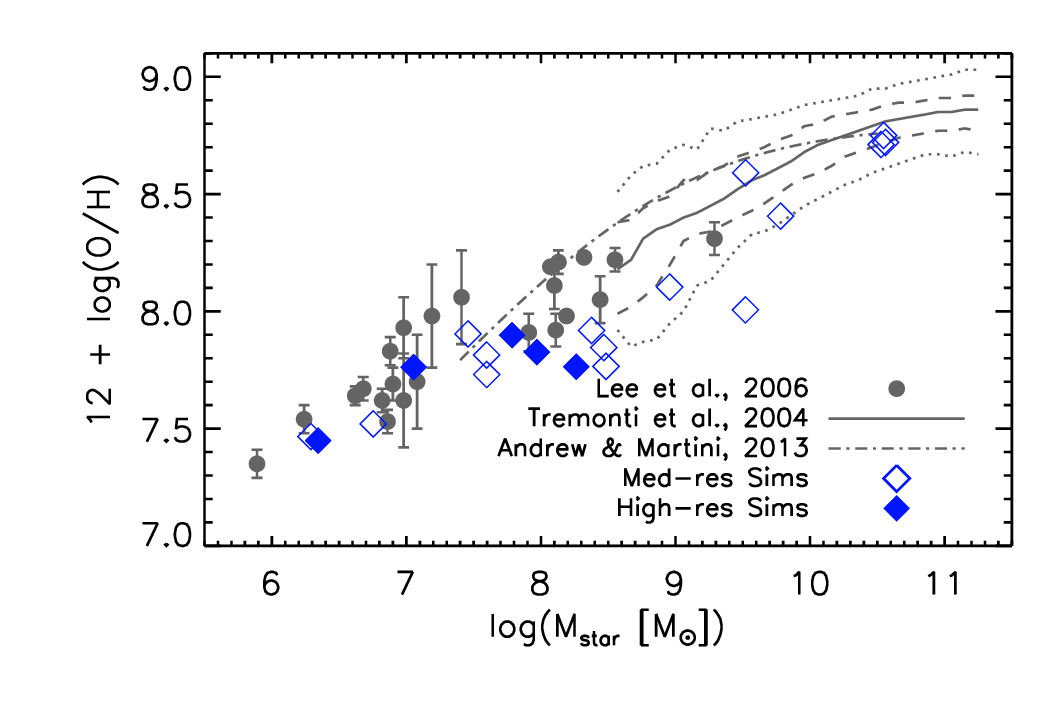}
\end{center}
\caption[Mass-Metallicitiy Relation]
{ 
The mass-metallicity relation for the simulated galaxies (blue diamonds) compared to observed galaxies.
Here the stellar masses of the simulations were calculated from mock-photometric observations.
Filled diamonds represent higher resolution simulations than the empty diamonds.
Grey circles represent individual observed galaxies from \citet{Lee06}.
Lines show fits from \citet{AndrewsMartini2013} (dot-dashed line) and  \citet{tremonti04} (the solid line is the median of the galaxies in bins of 0.1 dex in stellar mass, the dashed line is the contour enclosing 68\% of the galaxies, and the dotted line is the contour enclosing 95\% of the galaxies).
All lines from \citet{tremonti04}  were shifted down by 0.26 dex, as the method used in that paper produces systematical higher oxygen abundances  \citep{Erb06}.
}
\label{fig:mz}
\end{figure}

Figure~\ref{fig:mz} shows our simulated galaxies in relation to the
observed redshift zero mass-metallicity relationship from \citet{Lee06}, \citet{tremonti04} and \citet{AndrewsMartini2013}\footnote{Note that the method used in \citet{AndrewsMartini2013} differs from that of the other two observed samples.
In \citet{AndrewsMartini2013}, the direct method was used to measure the metallicity from stacked spectra of SDSS galaxies, whereas in \citet{Lee06} and \citet{tremonti04} metallicities were measured from the flux ratios of strong lines for individual galaxies.}.
The \citet{tremonti04} metallicity data were lowered by 0.26 dex
in order to account for the offset in the metallicity calibration, as noted by \citet{Erb06}. 
Following \citet{Lee06}, stellar masses for the simulations are calculated from the K and B-band magnitudes.  
Gas particle oxygen abundances are weighted by the particle's star formation rate (i.e. probability of star formation) to mimic the measurement of metallicities in star forming regions for observed galaxies.  
The simulated galaxies with stellar masses between $10^8$ and $10^9 M_\odot$ have somewhat lower metallicites than the observed galaxies.  
Nevertheless, the simulated galaxies broadly follow a power law with [O/H]$\propto M_*^{0.3}$ from $M_*\approx 10^{6.5}-10^{10.5}M_\odot$, which is in good agreement with the observed galaxies over that wide range of masses.

Taken together, the stellar mass-halo mass, Tully-Fisher and mass-metallicity relations provide stringent constraints on the cumulative effects of gas inflows, gas outflows, and star formation.
The agreement of the simulations with these relations imply that such processes and their scalings with mass are plausibly represented in our simulations.
We note that while the energy per supernova (dESN) and star formation efficiency (c$^*$) were adjusted together to match the stellar mass-halo mass relation, no other tuning was done. 
Furthermore, the agreement between the medium and high-resolution runs suggests that our results are not strongly
dependent on numerical resolution, with the caveat that we only probe a factor of two in spatial resolution.  
In the remainder of this paper, we examine how the properties of the outflows themselves scale with halo mass.

\csznote{
\subsection{Instantaneous Ejecta Properties}
\begin{itemize}
\item Velocity
\item Temperature
\item Pretty Pic
\item metallicity?
\end{itemize}
}

\subsection{Baryon fractions in the disk and halo}\label{sec:baryfrac}

The fraction of halo baryons in stars varies strongly with halo mass.  
Dwarf galaxies are known to be considerably less efficient than $L^*$ galaxies at forming stars, i.e., the mass fraction of the expected halo baryon content of dwarf galaxies in the form of stars is much smaller than for $L^*$ galaxies.
The reason for this must be some combination of three factors: {\it preventive feedback,} where baryons are prevented from accreting either onto the halo or onto the disk from the halo; {\it ejective feedback,} where material enters into the disk but is ejected back into the surrounding halo or beyond; and lower {\it global star formation efficiency,} in which gas enters and remains in the disk but is less efficient at forming stars.  
In this section we quantify the relative importance of these processes for simulated galaxies.

\begin{figure}
\begin{center}
\includegraphics[width=0.5\textwidth]{\path 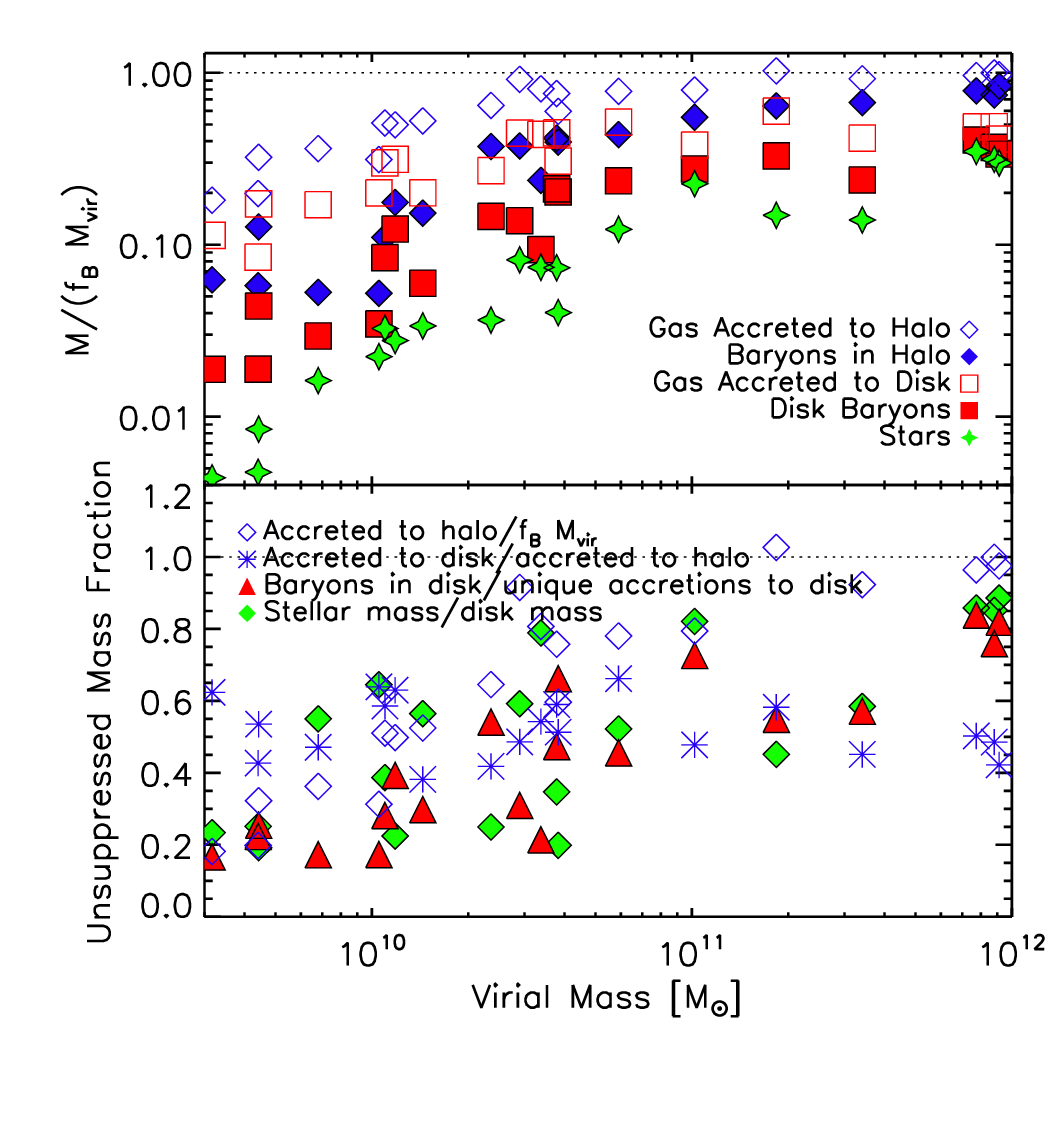}
\end{center}
\caption[Baryonic Mass Fraction]
{ 
The {\em top panel} shows fraction of the virial mass scaled by the cosmic baryon faction in different components of the galaxy.
In addition to showing the instantaneous galaxy properties at z = 0 (filled symbols), this plot indicates the total mass ever accreted into different components of the galaxy (empty symbols).
Blue diamonds represent all baryons ever accreted onto the halo (empty blue diamonds) and baryons presently within the halo (filled blue diamonds).
Red squares represent all the baryons ever accreted onto the disk (empty red squares) and all baryons presently part of the disk (filled red squares) in the form of cold, dense gas ($\rho \geq 0.1$ amu/cc and $T \leq 1.2 \times 10^4$) or stars.
Green stars represent the present stellar mass, with values taken {\bf directly} from the simulation.
The {\em bottom panel} shows relative efficiency of various methods at suppressing star formation.
Blue diamonds show the fraction of the cosmic baryons accreted to the halo, an indication of the importance of halo preventative feedback, while blue asterisks show the fraction of halo gas accreted onto the disk, a result of disk preventative feedback.
Red triangles show the fraction of baryons accreted to the disk that are still there at present day, which is a function of the ejective feedback.
Green diamonds (mass fraction of the disk in the form of stars) show the relative efficiencies of the galaxies in forming stars.
}
\label{fig:baryfrac}
\end{figure}

The top panel of Figure~\ref{fig:baryfrac} shows the redshift zero baryonic fraction of our simulated galaxies as a function of the expected cosmic halo baryon mass (i.e. the baryon fraction, $f_b$, times the halo mass).
The filled red squares show the $z=0$ fraction of baryons in the disk (cold gas plus stars), while the filled blue diamonds show the fraction of baryons in the halo.  
The corresponding open symbols show the total mass ever accreted into each component since a redshift of three (counting a given particle only once, even if accreted multiple times).  
Finally, the green stars indicate the present stellar baryon fraction.  

The bottom panel of Figure~\ref{fig:baryfrac} further summarizes the relative importance of preventative feedback, ejective feedback, and lower global star formation efficiencies in suppressing star formation.
In this panel, the fraction of the expected baryonic halo content that is accreted onto the halo is shown by blue diamonds, and the fraction of halo gas accreted onto the disk is shown by blue asterisks respectively.
The fraction of accreted baryons that remain in the disk at z = 0 is shown by red triangles and the stellar fraction of the disk is shown by green diamonds.
As can be seen from the top panel of Figure~\ref{fig:baryfrac}, all baryon fractions increase with halo mass.  
The stellar baryonic fraction goes from $\la 1$\% for $M_{\rm halo}\la 10^{10}M_\odot$ to about 20\% for $L^*$ halos.  

Preventive feedback manifests in two distinct forms in the plot, as shown by the open symbols in the top panel.  
The first is halo preventive feedback, in which the halo never receives its cosmic share of baryons.  
This is seen as the difference between the dotted line at unity and the open blue symbols.  
Next is disk preventive feedback, whereby the disk does not receive all the baryons accreted onto the halo.
This is quantified by the difference between the open blue symbols and open red symbols.

Addressing first halo preventative feedback, at virial masses below a few times $10^{10} \Msun$ the halos have reduced amounts of material ever accreted onto the halo.  
This halo preventative feedback is also evident in the bottom panel where the open blue diamonds mark the fraction of the expected baryonic content ever accreted to the halo.
The existence of halo preventative feedback for virial masses less than couple times $10^{10} \Msun$ is expected from the impact of the cosmic UV background, which reduces baryon fractions for galaxies with $M_{vir} \lesssim 10^{9.8} \Msun$ \citep{Gnedin00,Hoeft2006,Okamoto2008}.  
There is some uncertainty in this so-called filtering mass, so potentially our simulations are consistent with simple photoionization suppression, especially since only gas that was in the galaxy at z = 3 or later is included.
However, a distinct effect could owe to some other form of preventive feedback \citep[e.g.][]{Mo2002} such as heating by wind energy~\citep{Oppenheimer10,vandeVoort10}.  
At halo masses $>\sim 3\times 10^{10} \Msun$, halos generally accrete their fair share of baryons.

Disk preventive feedback, seen by the difference between the open blue diamonds and open red squares in the upper panel and the blue asterisks in the bottom panel, is typically around a factor of two.
This has only a mild dependence on halo mass, indicating that once baryons are inside the halo, an approximately uniform fraction of them reach the disk.  
This trend clearly cannot be extrapolated to masses above those considered here, since halos well above $10^{12} \Msun$ generally have substantial hot gaseous halos but have little gas in their disks (and typically do not have disks at all).  
These simulations are also missing possible sources of preventative feedback, namely heating from AGN and ionizing radiation from the galaxies' stars, that could possibly introduce a mass trend within star forming galaxies \citep{Kannan2013}.
Nevertheless, our simulations here indicate that disk preventive feedback is not strongly dependent on halo mass.

We now consider ejective feedback, which can be seen as the difference between the open and filled points of the same color in the top panel. 
For instance, the halo has accreted a baryon fraction indicated by the open blue diamonds, but currently only contains the fraction indicated by the filled
blue diamonds, so the difference must have been ejected from the halo.  
Similarly, the difference between the open and filled red squares quantify disk ejection. 
Disk ejection is also apparent from the red triangles in the bottom panel, which mark the fraction of mass accreted to the disk that remains in the disk at z = 0.

In our simulations, there is only one energetic process included that can counteract gravity, eject material, and produce galactic winds, namely supernova feedback.  
Hence ejective feedback quantifies how much material supernova feedback has removed from the disk and the halo.
Note that this does not necessarily mean that all ejection consists of gas directly heated by the supernova, as there could be some entrainment, pushing, or heating of the surrounding gas.  
Ultimately, though, the energy source must have been the supernovae.
As can be seen from the figure, ejective processes have a strong dependence on halo mass.  
Low-mass disks can eject the vast majority of accreted material, whereas for $L^*$ galaxies the disk ejection is much less.  
We further explore this mass dependency in \S\ref{sec:massloading}.
\csznote{
Halo ejection follows a similar trend, but interestingly the amount of halo ejection {\it relative to} disk ejection does not increase markedly to low masses.
This would be expected, for instance, if the effectiveness of ejective feedback was primarily determined by the halo potential well, as in the classic scenario of \citet{Dekel86} and many subsequent works.  
Our results are instead more consistent with the results of \citet{Oppenheimer08} who found that the kinematics of outflowing material is not well described as responding to gravity, but rather to hydrodynamic forces as the outflowing gas moves through ambient material.
}

Finally, we consider the global star formation efficiency, which is quantified by the difference between the solid red squares and the green stars in the top panel and by the green diamonds in the bottom panel.
Again, this shows a significant trend with halo mass, such that low-mass galaxies are less efficient at turning their disk baryons into stars.  
In observed galaxies this lower global star formation efficiency is evident in the higher gas fractions of dwarf galaxies.
The reason is a combination of at least two effects.
The first is that low-mass galaxies tend to be more diffuse, with the lowest mass systems often being bulgeless and irregular, and hence the star formation rate, which is superlinearly dependent on the local density, is reduced for a given amount of mass.  
Second, low-mass galaxies have lower metallicity, which reduces the amount of dust shielding, and consequentially, the $H_2$ fraction.
As a result, the dwarf galaxies have lower rates of conversion of disk gas into stars.
Overall, inefficiency in global star formation seem to be of comparable strength to ejective effects across all masses.

\csznote{
\begin{figure}
\begin{center}
\includegraphics[width=0.5\textwidth]{\path hgal_vm_eject_expell_fsuppress.eps}
\end{center}
\caption[Mass retained in disk]
{
Relative efficiency of various methods in suppressing star formation.
Blue diamonds show the fraction of the cosmic baryons accreted to the halo, an indication of the importance of halo preventative feedback, while blue asterisks show the fraction of halo gas accreted onto the disk, a result of disk preventative feedback.
Red triangles show the fraction of baryons accreted to the disk that are still there at present day, which is a function of the ejective feedback.
Green diamonds (mass fraction of the disk in the form of stars) show the relative efficiencies of the galaxies in forming stars.
}
\label{fig:supressfrac}
\end{figure}
}

\csznote{
These findings as to the relative importance of the preventative feedback, ejective feedback, and star formation efficiency are summarized in Figure ~\ref{fig:supressfrac}.
In this figure, the effect of preventative feedback is shown by the blue points: diamonds indicate the fraction of the cosmic baryons accreted to the halo (the same as in Figure~\ref{fig:baryfrac}) while asterisks show the fraction of gas accreted to the halo that is then accreted to the disk.
As previously discussed, halo preventative feedback is important at dwarf mass scales and disk preventive feedback is largely constant at 50\% across halo mass.
The importance of ejective feedback can be seen by comparing the current mass of baryons in the disk to the mass ever accreted onto the disk (red triangles).
Similarly, variations in star formation efficiency are apparent by comparing the stellar mass to the disk baryonic mass (green triangles).
Both ejective feedback and varying star formation efficiency reduce the stellar mass in lower mass halos and they are of similar importance across the mass range analyzed.
}

\begin{figure}
\begin{center}
\includegraphics[width=0.5\textwidth]{\path 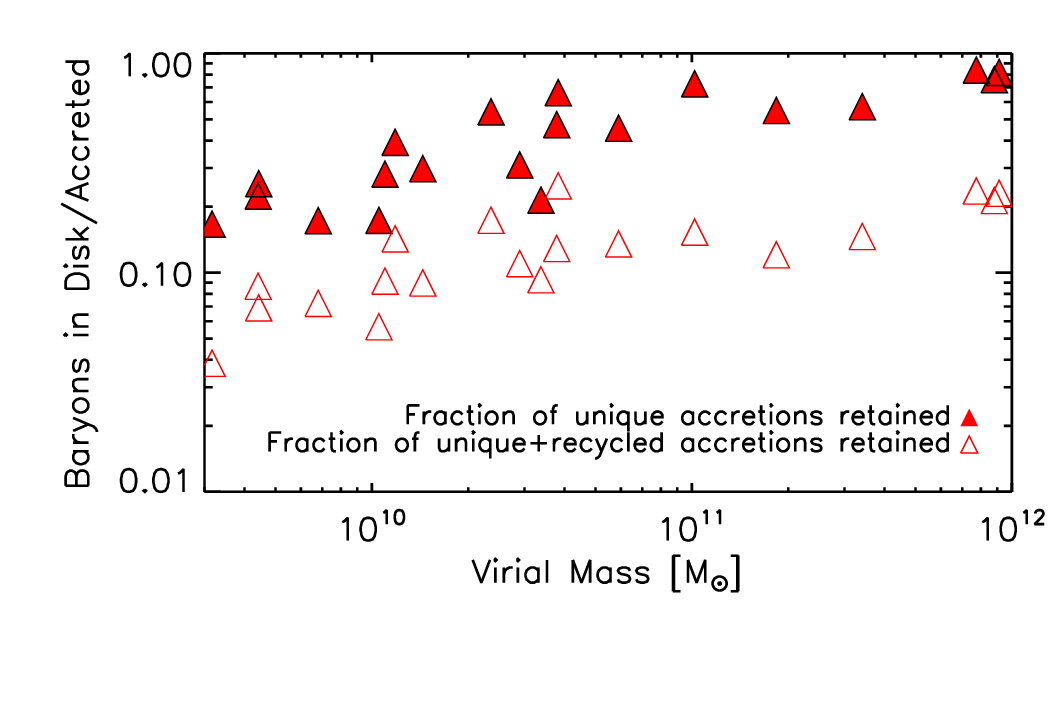}
\end{center}
\caption[Mass retained in disk]
{ 
The mass of disk baryons at redshift zero as a fraction of the total mass accreted onto the disk.
For the filled triangles, the denominator is the total mass of unique gas particles accreted.
For the empty triangles, the denominator includes multiple accretions of the same gas particle.
According to both definitions, the fraction of accreted baryons retained in the disk increases by about a factor of four across the halo range covered.
}
\label{fig:fracaccr}
\end{figure}

So far we have only considered unique accretion events, which specifically neglects recycling of previously ejected material back into the disk.  
To examine this we compare the disk baryon fractions relative to the total accreted and reaccreted fraction onto the disk in Figure~\ref{fig:fracaccr}.
In the case of the solid red triangles, we consider only unique accretions (i.e. not including recycling); this is equivalent to red triangles in the bottom panel of Figure~\ref{fig:baryfrac}. 
In the case of the empty red triangles though, we consider all accretions, including recycled material.

Several trends are evident.  
First, the fraction of both unique and total accreted mass that ends up in the disk has a strong dependence on halo mass.  
In the highest mass galaxies about 80\% of the baryons that are accreted once ended up in the disk, while in the lowest masses ones it is $\sim 4\times$ smaller.  
The {\it total} accretion, in all cases is also about $4\times$ smaller for unique events than for re-accreted events.  
That is, typically there is $3\times$ greater recycled accretion than first-time accretion.  
This number is roughly independent of mass, showing that the amount of recycling relative to first-time accretion is not strongly mass-dependent.  
We explore this topic in greater depth in \S~\ref{sec:cycling}.

\begin{figure}
\begin{center}
\includegraphics[width=0.5\textwidth]{\path 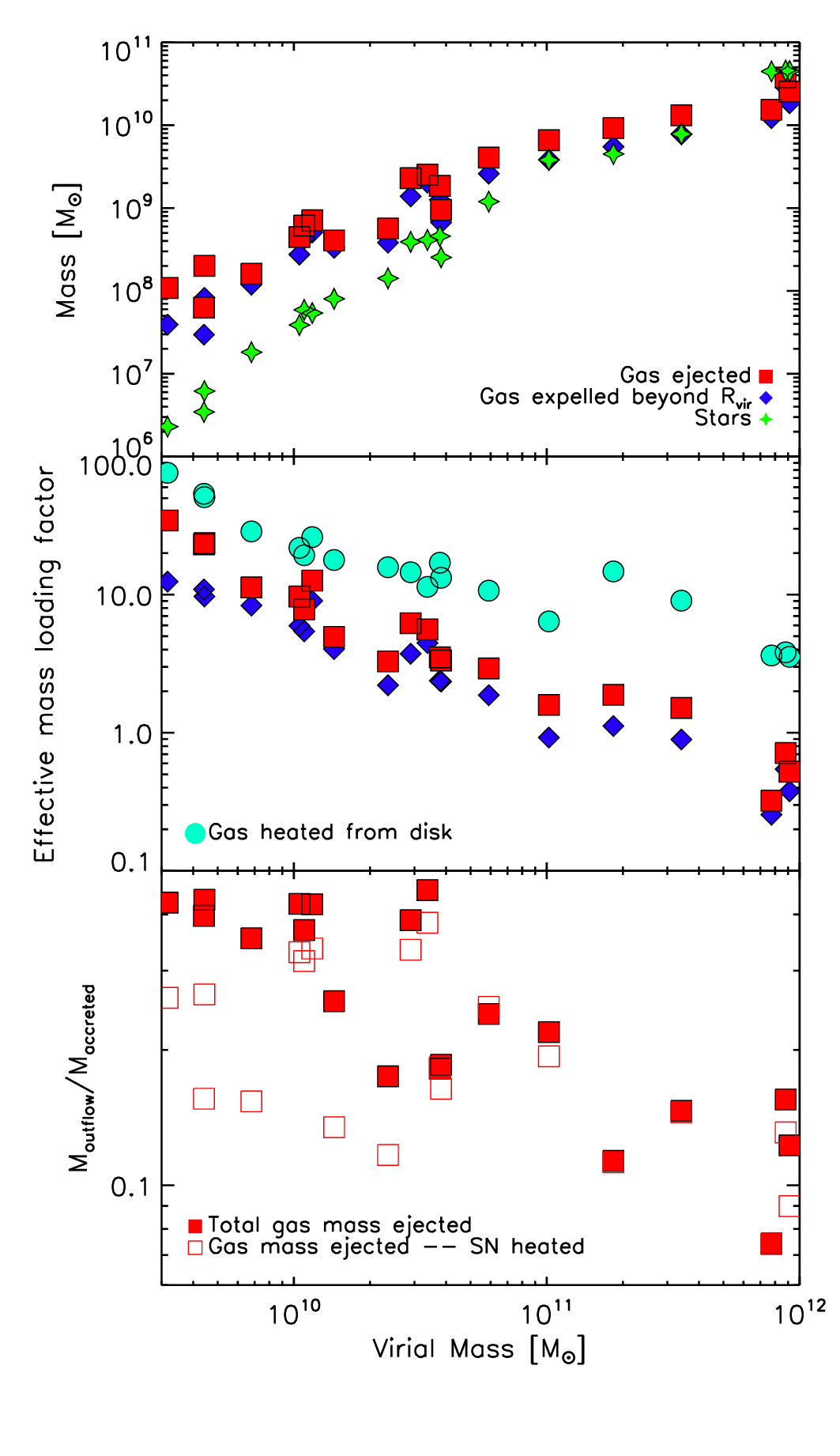}
\end{center}
\caption[Total gas mass lost from galactic disk over the histories of the galaxies]
{ 
In all plots, red squares represent the gas ejected from the disk.
Top plot: total gas mass ejected over the histories of the galaxies as a function of their virial mass.
Here the mass of ejected gas which eventually crosses the virial radius (blue diamonds) and the stellar mass of the galaxies (green stars) are shown for comparison.
The mass of the outflows is significant compared to the stellar mass and increases with galaxy mass.
Middle plot: total outflow mass divided by the stellar mass formed over the histories of the galaxies (the effective mass loading factor) as a function of the galaxy's redshift zero virial mass.
In addition to the gas ejected from the disk or expelled beyond the virial radius (blue diamonds), we also show all the gas that became too hot or rarefied to be considered part of the disk (filled green circles).
The effective mass loading factors display strong power law dependencies with mass.
Bottom plot: mass fraction of gas ever accreted onto the disk of the galaxy that is ejected from the galaxy.
Empty squares show the ejected gas particles that supernova energy was directly transferred to during the simulation, as opposed to gas particles that were entrained.
As predicted from the lower baryon fractions of dwarf galaxies, higher fractions of gas accreted onto the disk are later ejected from less massive galaxies.
Lower mass galaxies also lose greater fractions of their gas through entrainment rather than direct heating.
}
\label{fig:massloss}
\end{figure}

\subsection{Quantifying Ejected Material}\label{sec:totaleject}

\csznote{
\begin{figure}
\begin{center}
\includegraphics[width=0.5\textwidth]{\path hgal_vm_frac_gas_eject_expell_mass.eps}
\end{center}
\caption[Fractional mass lost from galactic disk over the histories of the galaxies]
{ 
Mass fraction of gas ever in the disk of the galaxy that is ejected (red) or expelled (blue) from the galaxy.
The difference between the fractions of ejected and expelled gas mass is remarkable.
At some point during the galaxies' history, most gas in the disk will be ejected because of supernova feedback.
However, less than half will ever escape beyond the virial radius.
For the expelled gas there is a slight mass trend with the less massive galaxies loosing more than half the gas ever in the disk.
The mass trend is less strong for the ejected material, probably because the increased potential of the halo has less of an effect on the likelihood of a particle escaping the disk than on the likelihood it will escape the halo.
{\bf CHARLOTTE: Show a line with unity slope somewhere on this plot.}
}
\label{fig:fracmassloss}
\end{figure}
}

We now examine in more detail the role of ejective feedback in regulating the baryonic content for galaxies of different masses.
Specifically, we quantify the mass of outflows that galaxies experience as a function of their virial mass.  
We quantify outflows in several ways: (1) the gas that becomes too hot or rarified to be considered part of the disk; (2) the gas that becomes energetic enough to dynamically escape the baryonic  disk (``ejected"); and (3) the gas not only ejected from the disk but also expelled from the entire halo.  
Each of these latter categories is a subset of the former.  
We can also subdivide any one of those categories between the gas particles which directly had supernova energy transferred to them and those that were entrained by other affected gas.
Note that, with regards to Figure~\ref{fig:fracaccr}, all the outflow masses should be compared to the total accretion including recycling as opposed to unique accretion, since outflows also must eject recycled material.

Figure~\ref{fig:massloss}, top panel, shows the total amounts of gas ejected from the disk (red squares) and expelled from the halo (blue diamonds) since a redshift of 3 as a function of the virial mass.  
For comparison, the mass in stars formed over the history of the galaxy is also shown (green stars).  
For galaxies with $M_{\rm halo}\la 10^{11} \Msun$ the amount of ejected material exceeds the stellar mass and at the lowest masses it does so by an order of magnitude.  
This leads to the conclusion that, particularly for smaller halos, the visually-dominant stellar component is merely a ``trace" amount of mass leftover from the small imbalance between accretion and outflows.

The middle panel of Figure~\ref{fig:massloss} quantifies the mass loss more precisely via the {\it effective mass loading factor} $\tilde\eta$, defined as the amount of mass driven out by feedback relative to the amount of stellar mass formed.  
This is different than the more canonical mass loading factor $\eta$, which is the instantaneous mass loss rate relative to the current SFR.  
The effective mass loading factor is in some sense a time-average of $\eta$ over a galaxy's life.  
We note that, owing to our limited snapshot time resolution of $\sim 100$~Myr, $\tilde\eta$ is actually a lower limit on the true value, since there could be recycling happening on smaller timescales that would add to the total outflow mass.

The effective mass loading factor, $\tilde\eta$, displays the trends expected from the top panel.
For our most massive halos, $\tilde\eta$ (either for material ejected from the disk or expelled beyond the virial radius) is comparable or less than unity, showing that galaxies retain in stars roughly as much material as they expel.  
At small masses, however, $\tilde\eta$ for ejected material can reach up to $\sim 30$, and for the subset of the ejecta that is expelled beyond the virial radius, $\tilde\eta \sim 10$.  
This relationship demonstrates the extent to which less massive galaxies are more efficient at removing material from the galaxy through stellar feedback; we will quantify this further in the next section.

We also show the total mass of gas heated by feedback relative to the stellar mass as the green circles.
Such gas is heated enough to no longer be considered part of the atomic and molecular ISM but may never have enough kinetic energy to be considered ``ejected" from the disk.
Therefore, this group includes gas which is briefly heated by supernovae but quickly cools back onto the disk.
This comprises a significant amount of material and shows the impact supernovae can have on the gas available for star formation, even when not removing the gas from the disk for great lengths of time.

At large masses, most of the disk baryons are in stars, and so $\tilde\eta$ approximately reflects the amount of ejection relative to the disk baryons.  
But at small masses this is not true, because the lower star formation efficiency results in large gas reservoirs that far exceed the stellar component.  
Hence a third quantitative view of outflows is to compare the outflow mass to the total amount of baryons ever accreted onto the disk.

The bottom panel of Figure~\ref{fig:massloss} shows the fraction of ejected material relative to the total accreted mass (including recycled accretion).  
The open squares show the ejected material that was directly heated by SNe, while the filled squares show the total amount of ejected material.  
As can be anticipated from the reduced baryonic fraction of dwarf galaxies, higher fractions of accreted material are lost as the virial mass decreases.  However, the trend is not as strong or as clean as those for $\tilde\eta$ because the lower star formation efficiencies of smaller mass galaxies results in much of the accretion going into the gas reservoir rather than being ejected.  

The effect of entrainment can be seen by comparing the open and filled symbols in the bottom panel. 
At high masses, there is little entrainment; most of the gas particles either ejected or expelled had energy transferred to them directly from supernovae at some point in their history.  
In contrast, there is substantial entrainment of gas in dwarf galaxies, in some cases as much as doubling the mass of the SN-heated material.


\begin{figure}
\begin{center}
\includegraphics[width=0.5\textwidth]{\path 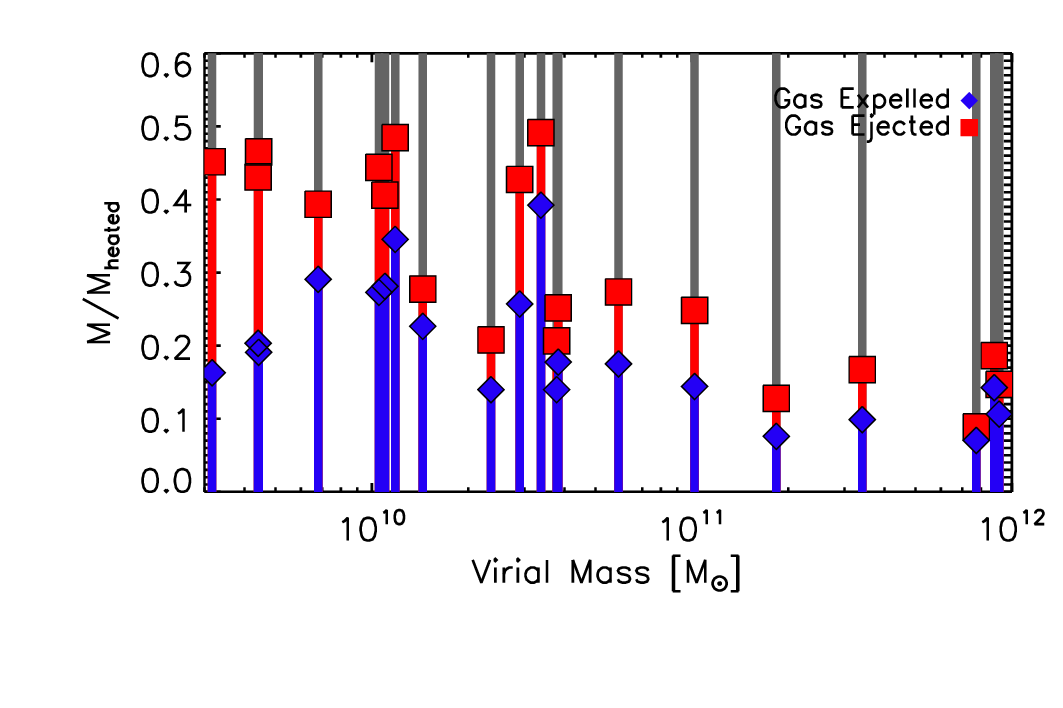}
\end{center}
\caption[Fraction of ejected gas that was expelled from the galaxy]
{ 
Fraction of gas heated sufficiently to be removed from the disk (M$_{heated}$) that was either ejected from the disk (red) or expelled beyond the virial radius (blue). 
The fraction of M$_{heated}$ that is ejected strongly decreases with virial mass.  
This trend can be accounted for by the relatively larger disk masses in more massive halos.
The fraction of M$_{heated}$ that is expelled from the halo shows a similar but far weaker trend with mass.
Also evident is the decreasing fraction of ejected particles later expelled from the halo.
}
\label{fig:frac_ejected_expelled}
\end{figure}

In order to better analyze the division between all gas that was heated by supernovae and the subsection that was ejected from the disk or expelled beyond the virial radius, we show the relative fractions of each for the different halo masses (Figure~\ref{fig:frac_ejected_expelled}).  
In this figure each gray bar represents the total mass of gas that became too hot or rarefied to be considered part of the disk (corresponding to the green circles in the middle panel of Figure~\ref{fig:massloss}), while the red shows the fraction of that gas that was ejected from the disk and the blue the fraction that was expelled beyond the virial radius.
The fraction of the heated gas that was ejected (and, to a lesser extent, expelled beyond the virial radius) decreases with increasing halo mass.  
This trend demonstrates the greater difficulty in gas becoming energetic enough to dynamically leave the disk or halo of more massive galaxies.
Furthermore, the fraction of ejected particles that were later expelled beyond the virial radius also decreases with halo mass.  
In the lowest mass galaxies, less than half of the ejected gas eventually leaves the halo whereas in the higher mass galaxies almost all of the ejected gas does.
This trend in the fraction of ejected gas later expelled from the halo is the result of the relatively larger disks in more massive halos -- the larger mass the disk is relative to the halo, the more likely it is that a particle that becomes dynamically unbound from the disk will exit the halo.


\subsection{Mass loading factor evolution}\label{sec:massloading}

\begin{figure}
\begin{center}
\includegraphics[width=0.5\textwidth]{\path 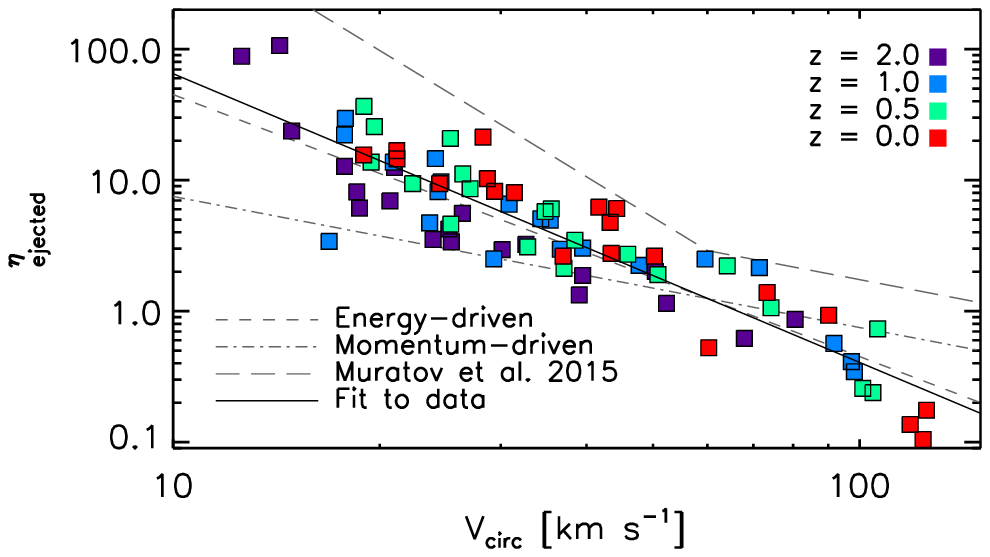}
\end{center}
\caption[Average massloading as a function of galaxy stellar mass]
{
Total gas mass ejected divided by the stellar mass formed within 1 Gyr time bins as a function of halo circular mass.
The mass loading factors at different redshifts are indicated by different color points.
A power law fit to all the data points, shown as the black line, results in an exponent of -2.2.
For comparison, the dashed and dot-dashed lines show an energy driven scaling ($v_{\rm circ}^{-2}$) and a momentum driven scaling ($v_{\rm circ}^{-1}$), respectively (normalized to the data at $v_{\rm circ}$ = 60 km/s).
Additionally, the redshift zero fit to the mass loading factor given in \citet{Muratov2015} is included as a long dashed line.
Note that differences between how the ejecta was selected here and in \citet{Muratov2015} can account for differences in the offset but not in the scaling.
}
\label{fig:mass_massloading}
\end{figure}

We now examine the mass loading factor, $\eta$, as a function of redshift and halo mass.  
We obtain a more ``instantaneous" mass loading factor by taking the total mass lost only in the past
gigayear at any given redshift and comparing it to the mass of stars formed
over the same time period.  
While this is not truly instantaneous, it allows us to examine how this quantity has varied with cosmic time.  
We also examine $\eta$ as a function of circular velocity, as that is more straightforwardly relatable to recent models of the physics of outflow driving.
In this case, ``circular velocity" refers to the circular velocity calculated at the virial radius from the total halo mass.

Figure~\ref{fig:mass_massloading} shows the mass loading factors for the ejected material at four different redshifts, against the corresponding circular velocities at those redshifts.
There is no discernible trend with redshift in either shape or amplitude.  
Fitting all the values together, we obtain a power-law fit of $\eta = \eta_0 v_{\rm circ}^{-2.2}$  for
ejected material, roughly consistent with energy-driven winds \citep{Chevalier1985}.  
\csznote{
The best-fit slope is slightly shallower for the expelled case, because the baryonic disk mass decreases more rapidly with circular velocity than the total halo mass.  The scatter is also less in the
ejected case relative to the expelled, which is perhaps unsurprising
since the physical driver of the outflows is from the disk.  Overall,
this scaling is consistent with the notion that the outflows in our
simulations are being driven by supernova energy.
}

Cosmological simulations often have to assume a mass loading factor since they lack the resolution to directly generate outflows.  
The energy-driven scalings we predict are consistent with those assumed in state-of-the-art simulations that match observed properties of galaxies and the circum-galactic medium \citep{Dave2013,Ford13b,Vogelsberger2014,Genel2014}.  
They are, however, somewhat different than that predicted by the Feedback in Realistic Galaxies (FIRE) suite of zoom simulations that also self-consistently drive outflows; the FIRE simulations find a shallower dependence for $v_{\rm circ}>60$~km/s, and a steeper dependence for smaller systems~\citep{Muratov2015}.
The amplitude is somewhat lower than typically assumed in cosmological runs; we predict that $\eta$ is unity for $v_{\rm circ}\approx 100$~km/s (i.e. $M_{\rm halo}\sim 10^{11}\Msun$), whereas such simulations typically assume unity mass loading at $M_{\rm halo}\sim 10^{12}\Msun$, corresponding to mass loading factors a factor of two smaller than typically assumed \citep[e.g.][]{Dave2013}.  
However, these mass loading factors are only calculated for the outflowing gas that is sufficiently energetic to be classified as ``ejected."
If all the gas identified as leaving the disk is included, the scaling of the mass loading increases to values closer to those assumed in most cosmological simulations.

\subsection{Velocities of outflows}\label{sec:vel}
\begin{figure}
\begin{center}
\includegraphics[width=0.5\textwidth]{\path 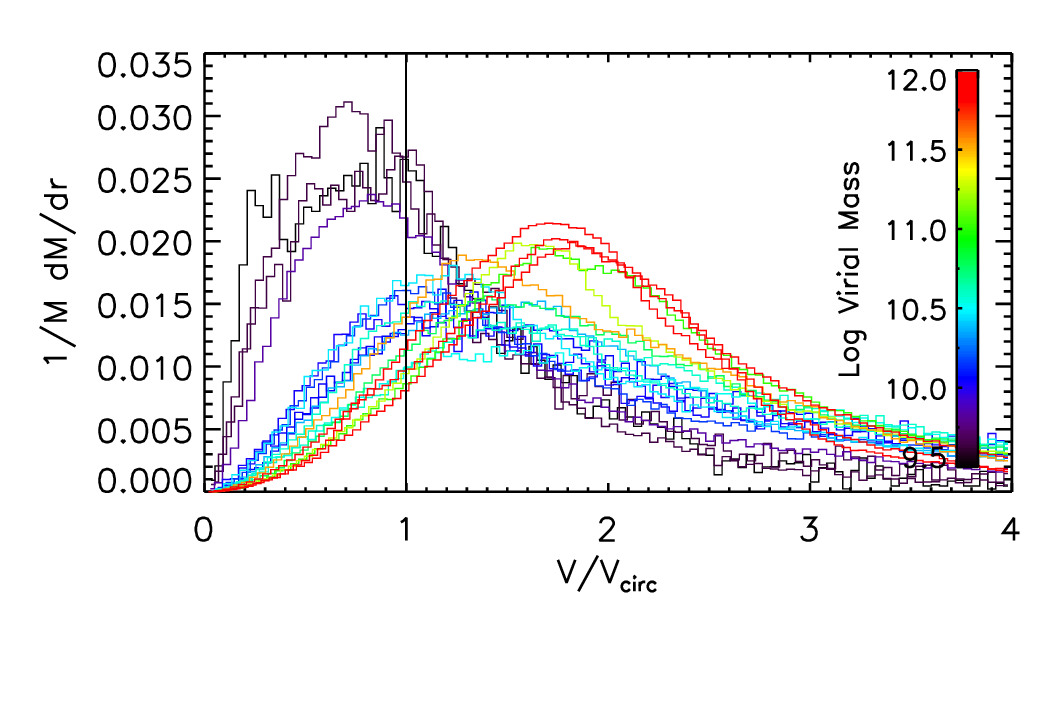} 
\end{center}
\caption[Velocity distribution of ejected material]
{ 
Normalized distribution of the velocities of the ejected material for individual galaxies.
The highest mass galaxies are shown in red whereas the lowest mass galaxies are shown in purple.
The velocities of the ejected material were determined at the step immediately following their removal from the disk.
The velocities of the ejected material are then scaled by the circular velocity defined for the halo potential at the time when the particle was ejected.
}
\label{fig:vejecta_distib}
\end{figure}

\begin{figure}
\begin{center}
\includegraphics[width=0.5\textwidth]{\path 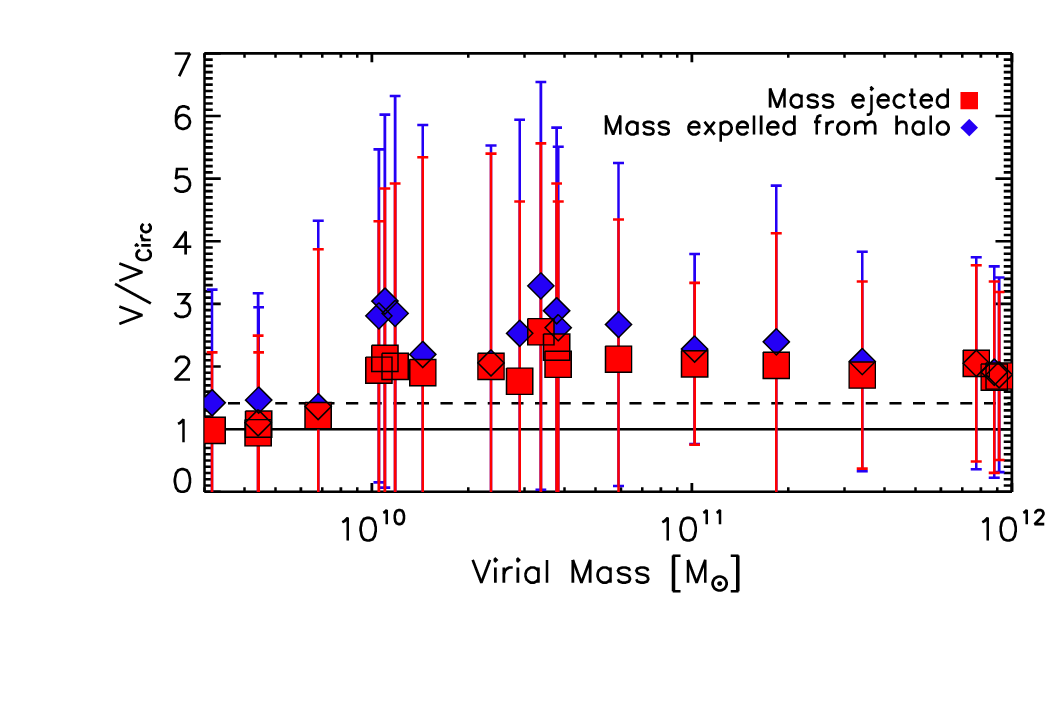} 
\end{center}
\caption[Velocity of ejected material]
{ 
Velocities of ejected and expelled material.
The median velocity of the material at the step immediately after ejection from the disk is shown as a function of virial mass.
These velocities are normalized by the circular velocity of the halo at the time of ejection.
This plot shows both the total ejected gas (red squares) and the subsection of that gas which is later expelled (blue diamonds).
The solid line marks the velocity equal to the circular velocity while the dashed line marks the escape velocity of the halo.
}
\label{fig:vejecta}
\end{figure}

Outflow velocities are generally observed to be proportional to the
circular velocity~\citep[though see \citealt{Steidel10}]{Martin05,Weiner09}, and cosmological simulations that employ kinetic feedback typically assume such a scaling.
Because our blastwave feedback depends on the local density and pressure, affected particles show a distribution of energies, temperatures, and velocities.  
We can thus compare these self-consistently generated velocities to those observed and employed in other models.

Figure~\ref{fig:vejecta_distib} shows the distribution of the velocities of ejected gas particles in the snapshot following their removal from the disk scaled by the circular velocity of the galaxy at the time of the outflow.  
Because of the spacing between snapshots (100 Myrs), particles are typically already 0.04 $R_{vir}$ from the center of the galaxy when their velocity is measured.
As such, particles may have slowed down during the period of time between when they actually left and when they are identified as having done so.
Nevertheless, this method allows us to make approximate measurements and to compare velocities across galaxies.
As in \S\ref{sec:massloading}, the circular velocity used here is the circular velocity calculated at the virial radius for the total mass of the halo.
Other definitions of rotational velocity will produce different scalings.  
For instance, the asymptotic velocities, $V_f$, used for the baryonic Tully Fisher relation (Figure~\ref{fig:btf}) are about 1.4 times greater than the circular velocities for the halo mass range here.
Normalizing the outflow velocities by $V_f$, therefore, would reduce the results shown by a factor of $\sim$0.7.
For the most massive galaxies, the median outflow velocities are close to twice the circular velocity at the time of the outflow.  
For the least massive galaxies, in contrast, it is clearly smaller.  
This lower velocity of gas ejected from dwarfs galaxies is consistent with their relatively smaller fractions of gas leaving the halo seen in Figure~\ref{fig:frac_ejected_expelled}.
Essentially, the relatively smaller disk-to-halo mass ratio of dwarf galaxies enables gas to be ejected at relatively smaller velocities.
However, these smaller velocities are not necessarily sufficient for the ejected material to escape the halo.

Figure~\ref{fig:vejecta} quantifies the velocity trend with halo mass and circular velocity in more detail.  
Here we compare the median velocities of the ejected material and the subset of that material that later escapes the halo at the snapshot following their removal from the disk as a function of the redshift zero virial mass.  
Red points show the velocity of gas ejected from the disk, while blue points show the velocity of the subset of that gas that will end up expelled from the halo.  
The ejection velocity increases with halo mass and median velocities are between one and two and a half times the circular velocity for all halo masses.
However, there is a huge range of velocities for any given galaxy, as indicated by the error bars.
Taking a closer look, there is a noticeable trend that intermediate-mass galaxies have the highest relative outflow velocities.
There is also a weak trend for material that later escapes the halo to have higher velocity, but there is not a strong correlation between the velocity at the timestep following when it exits the disk and its ability to escape the halo.

\csznote{
Interestingly, the velocities predicted in these simulations are considerably lower than what is typically assumed for outflows galaxies in cosmological simulations, which are generally close to the escape velocity or higher~\citep{Oppenheimer2006}.  
Such models justify this by comparing to observations of high-$z$ galaxies that show comparably rapidly outflowing material~\citep[e.g.][]{Steidel10}.
However, such observations may only be probing the ``tip of the (velocity) iceberg" of the outflow, whereas the bulk of the material likely has more modest velocities as predicted here.  
The very high outflow velocities (and also mass loading factors) assumed in those cosmological simulations are evidently not necessary to suppress star formation and metal growth in accord with data.
}

\subsection{Wind Recycling}\label{sec:cycling}

Previous theoretical work has shown that the recycling of gas through reaccretion back onto the disk is key to reproducing observed stellar masses\citep{Oppenheimer10,Henriques2013}, metal enrichment~\citep{Dave11b} and mass distributions \citep{Brook11b}.
In this section, we quantify re-accretion, or wind recycling, as a function of halo mass, as well as the distribution of recycling times.

\begin{figure}
\begin{center}
\includegraphics[width=0.48\textwidth]{\path 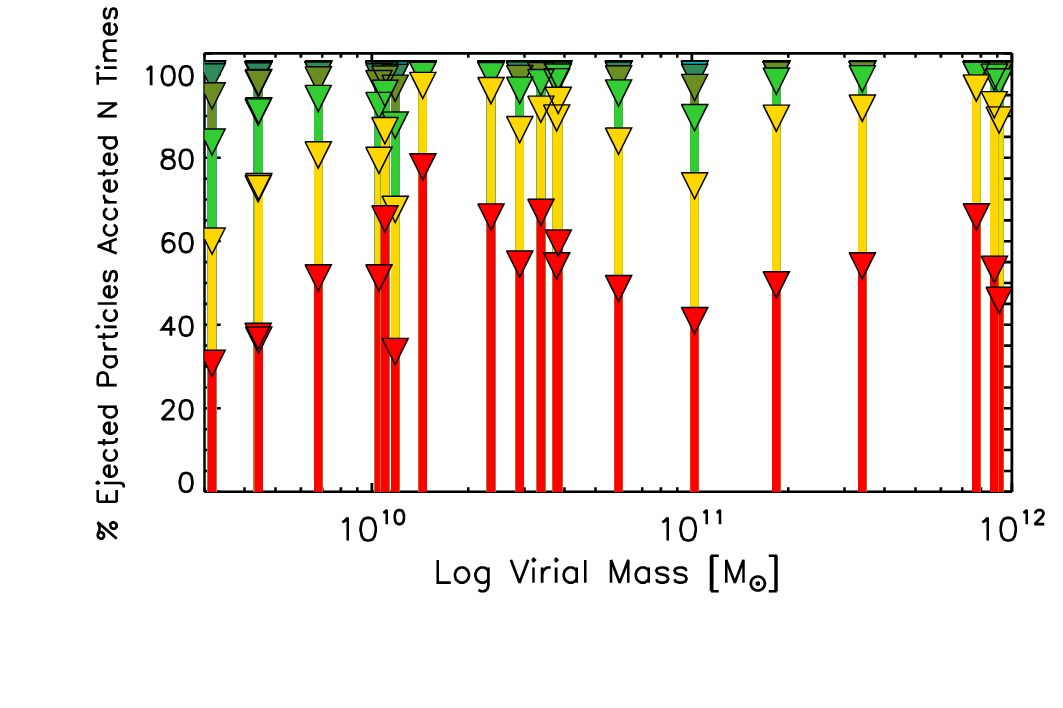}
\end{center}
\caption[Number of times a given particle is accreted onto the disk.]
{ 
Mass fraction of gas ever ejected that was accreted some number of times onto the disk as a function of the virial mass.
Each bar represents an individual galaxy.
The red marks the fraction of ejected particles never reaccreted during the history of the galaxy.
The yellow marks the fraction reaccreted once, the light green marks the fraction reaccreted twice and so on.
Between 20 and 70\% of ejected material is later reaccreted.
}
\label{fig:timesreaccrted}
\end{figure}

In Figure~\ref{fig:timesreaccrted} we show the fraction of particles that has been re-accreted some number of times after being
ejected from the disk.
The number of times particles are re-accreted reveals the importance of gas recycling to the baryon content of galaxies and the mass of the disks.
Across all galaxy masses, a significant fraction (20--70\%) of ejected mass is reaccreted onto the disk of the galaxy.
There is possibly a slight mass trend towards more massive galaxies experiencing less reaccretion of their ejected material.
As in the analysis of the fraction of ejected mass that later leaves the halo (Figure~\ref{fig:frac_ejected_expelled}), such a trend is likely the result of the varying disk-to-halo ratio.
Since the disk mass is a smaller fraction of the overall halo mass for dwarf galaxies, a greater fraction of the particles able to escape the disk potential are reaccreted because of the halo potential.

In contrast, we found no mass trend in the fraction of expelled material which is subsequently reaccreted.
For all galaxies, only about 20\% of the expelled material is reaccreted at least once.
As this process is largely controlled by the growth of the halo (i.e., expelled particles can be reaccreted once the potential well is deeper) it is reasonable for it to be independent from halo mass within a hierarchical growth framework.

\begin{figure}
\begin{center}
\includegraphics[width=0.5\textwidth]{\path 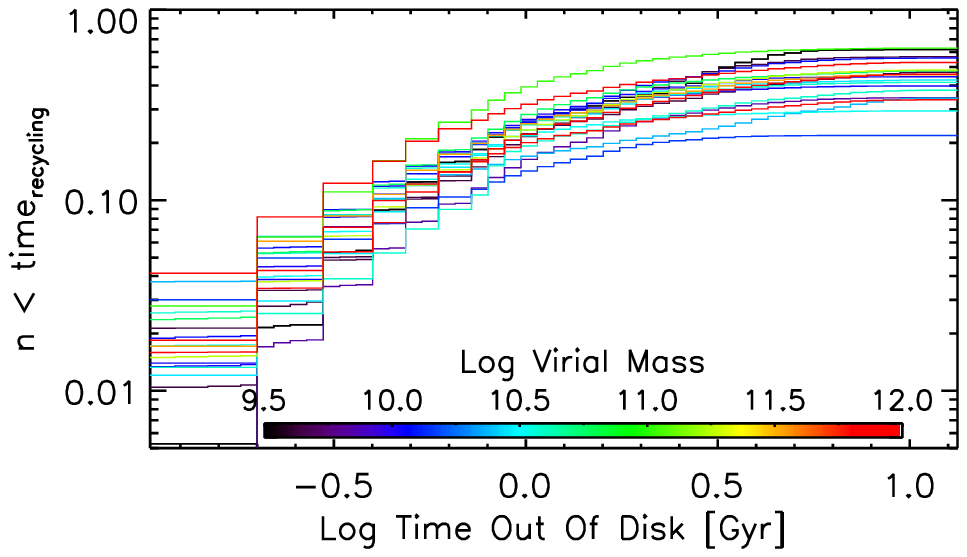} \\
\includegraphics[width=0.5\textwidth]{\path 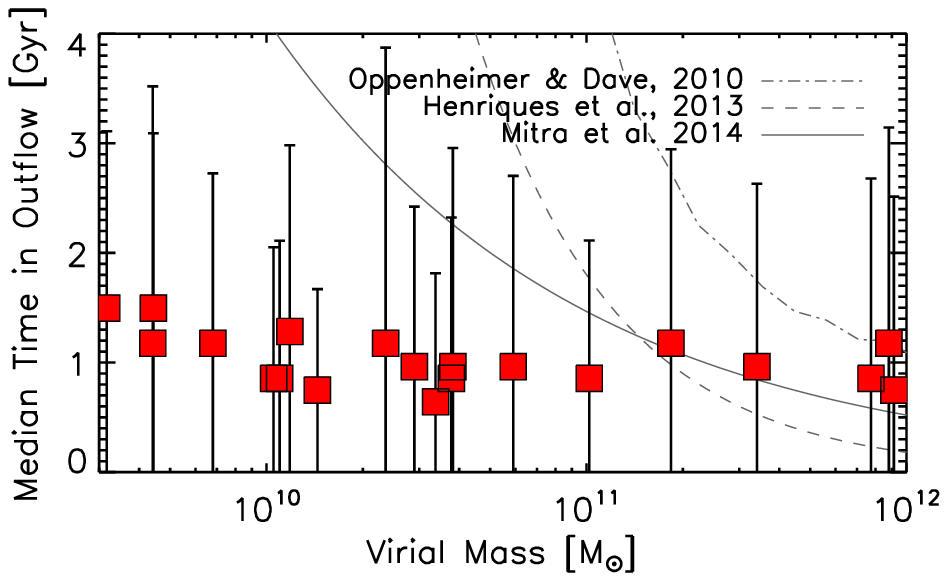} 
\end{center}
\caption[Amount of time spend out of the disk]
{ 
Amount of time before reaccretion of ejected particles to the disk.
Top: normalized cumulative histogram of the amount of time particles spend between their ejection and subsequent reaccretion onto the disk.
The colors represent the virial mass of the galaxies, with red being the largest mass halos and purple being the smallest mass halos.
Where these lines asymptote indicates the total fraction of gas ever reaccreted.
Bottom: median time for reaccretion of ejected particles as a function of halo mass with the error bars representing the standard deviation.
We find that on average particles take 1 Gyr to reaccrete and the recycling times have only a very weak dependency with halo mass.
This result is in contrast to the halo mass dependency found by \citet{Oppenheimer10} ($z$ = 0.5 from the preferred 'vzw' model), \citet{Henriques2013}, and \citet{Mitra2014}, shown here as the dot-dashed, dashed, and solid lines, respectively.
}
\label{fig:cycletime}
\end{figure}

A key parameter discussed in current galaxy formation models is the recycling time, i.e. the amount of time ejected gas spends outside the disk prior to re-accretion.  
Cosmological simulations predict that the recycling time for momentum-driven wind scales roughly inversely with the halo mass for moderate-sized star forming
galaxies~\citep{Oppenheimer08,Oppenheimer10}, and some semi-analytic models have found that similar scalings are best able to reproduce various observations, such as the stellar mass function~\citep{Henriques2013}.
Here we directly track the recycling time in our simulations as a function of halo mass.

Figure~\ref{fig:cycletime} shows cumulative histograms of the reaccretion time scales normalized by the total number of ejected particles.  
The recycling time distribution has a similar shape for all halos.  
Most gas particles are reaccreted on short time scales, with most reaccretion taking place over 500 Myrs.  
Since our particle tracking is limited by our 100 Myr snapshot time resolution, we are unable to track particles
that have recycling times less than 100 Myrs; this fraction is
likely not insubstantial.  Conversely, the recycling time distribution
also has a long tail, with some gas taking many gigayears to reaccrete.

In Figure~\ref{fig:cycletime}, we also show the median and standard
deviation of the reaccretion times as a function of virial mass.
This shows a very weak halo mass dependence, roughly $t_{\rm rec}\propto
M_{\rm halo}^{-0.1}$, with a typical reaccretion timescale of about 1 Gyr.  
This is in stark contrast to cosmological
simulations and semi-analytic models that seem to favor a strong
halo mass dependence.  Interestingly, the analytic equilibrium model
of \citet{Mitra2014} finds an optimal fit to the stellar mass and halo mass, star formation rate, and metallicity from 0 \textless z \textless 2  by
assuming a recycling time with a weak halo mass dependence of $t_{\rm
rec}\propto M_{\rm halo}^{-0.45}$.  This is still stronger than the
dependence our simulations predict, but is closer.

\subsection{Ejecta Properties}\label{sec:wherewhat}

Star formation-driven outflows have been shown to have a strong effect on the angular momentum distribution of both baryons and dark matter \citep{Brook11b}.
Outflows preferentially remove low-angular momentum central gas and, because of cloud-corona interactions, gas is re-accreted with higher angular momentum \citep{Marasco2012}.  
The effectiveness of this processes depends on where in the disk most outflow material is launched from and where it reaccretes to.
Here we examine these processes over a range of galaxy mass by tracking the location and angular momentum of particles at the time of ejection and reaccretion.

\begin{figure}
\begin{center}
\includegraphics[width=0.5\textwidth]{\path 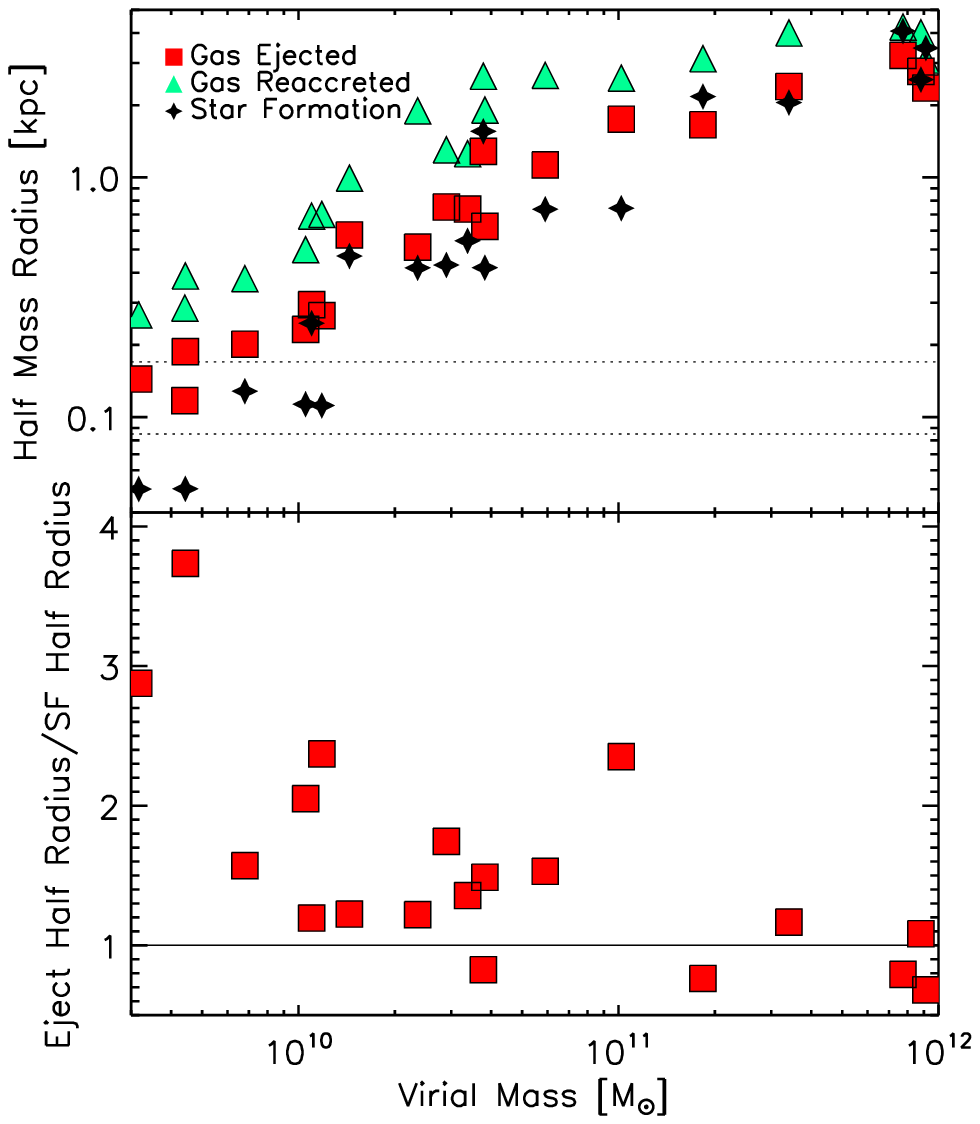}\\
\end{center}
\caption[Half-mass radius of ejected material]
{ 
Source of the ejected material as a function of virial mass.
The top panel shows the half-mass radius of the source of the ejected material (red squares), the reaccretion of the ejected material (green triangles) and the star formation (black stars). 
Here the half-mass radius is defined to be the radius from within which 50\% of the ejected material originates or is reaccreted to or within which 50\% of the star formation took place.
The dotted horizontal lines mark the softening lengths for both resolutions of galaxies.
The half-mass radii of the source of the ejecta, the location of reaccretion, and star formation increase with virial mass, as expected from the increasing size of the galaxies.
For all galaxies, gas is reaccreted to a significantly larger area than it is ejected from.
The bottom panel normalizes the half mass radius of the ejected material by that of the star formation.
The source of the ejected material roughly follows that of the star formation for galaxies above $10^{10} \Msun$.  For the four least massive galaxies, the half mass radii of star formation is within the softening length (87 pc), which limits our ability to draw conclusions as to the relative locations of star formation and outflows in these galaxies.
Nevertheless, it does appear that low mass galaxies eject gas from a broader region compared to their star formation.
}
\label{fig:reject}
\end{figure}

Figure~\ref{fig:reject} (top panel) shows the radius within which half of the
ejected mass originated and is reaccreted as a function of halo mass.  In our analysis,
the originating radius is defined by the location of the gas
particle at the snapshot immediately prior to being ejected and the reaccretion radius by the location of the gas particle at the snapshot following its reentry into the disk.    Similarly, the half-mass
radius of star formation is the radius within which half of the
z = 0 stellar mass was formed, which we obtain by tracking the
location of each star particle at the time of creation.

Both the star formation radius and the ejection radius increase
strongly with virial mass, roughly following a power law $R\propto
M_{\rm halo}^{1/4}$ which is somewhat shallower than the halo virial
scaling: $R\propto M_{\rm halo}^{1/3}$.  This slight difference is consistent with
the idea that lower-mass galaxies generally have a later Hubble
type, are more extended, and have more angular momentum.

In the bottom panel of Figure~\ref{fig:reject}, we normalized the
half-mass radius of the ejected material by the half-mass radius
of star formation.  The resulting plot shows that the outflowing material roughly follows the star formation, as expected.
There does appear to be a mass trend such that more massive galaxies have relatively more concentrated outflows.
However, this trend is largely set by the four lowest mass galaxies, whose star formation half mass radii are within the softening length, so it is unclear how robust this trend it.
It is possible, though, that lower levels of star formation 
Even for the more massive galaxies, though, the half-mass radius of ejected material is between 0.8
and 2.5 times the SF radius.  From the typically greater half-mass
radii of the ejecta, it is apparent that while the ejected material
is centrally concentrated, it is somewhat more dispersed than the star
formation. 

\begin{figure}
\begin{center}
\includegraphics[width=0.5\textwidth]{\path 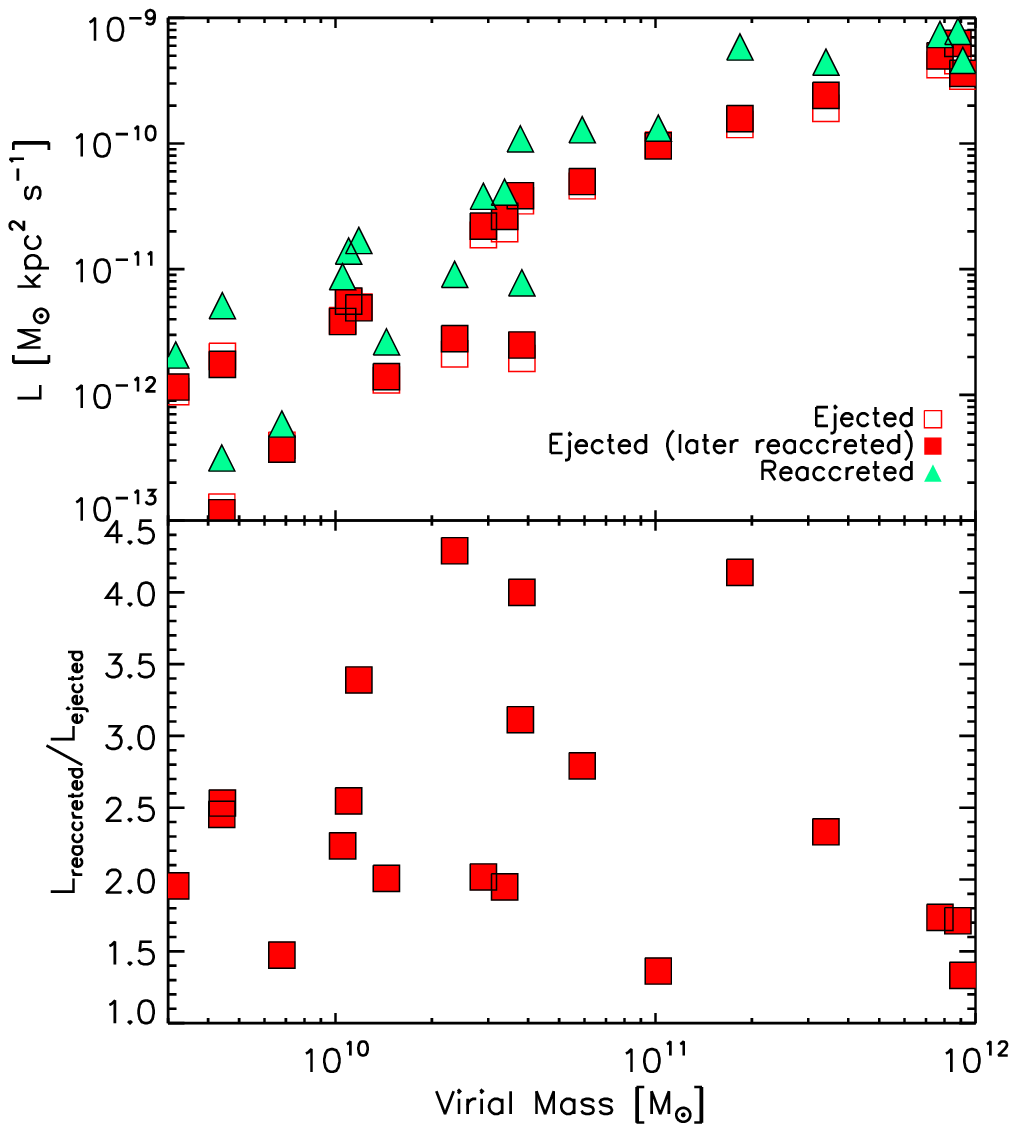}\\
\end{center}
\caption[Angular momentum of ejecta]
{ 
Median angular momentum of the ejected and reaccreted material as a function of virial mass.
The top panel shows the median angular momentum of all the ejected material (red empty squares) and the ejected material that is later reaccreted (red filled squares) at the snapshot immediately prior to it leaving the disk.
Green triangles indicate the median angular momentum of the reaccreted material at the first snapshot after it reenters the disk. 
The bottom panel plots the ratio between the median angular momentum of the material after reaccretion to the median angular momentum of it prior to its ejection as a function of virial mass.
On average, gas is reaccreted at significantly higher angular momenta, indicating that it is ``spun up" in the halo.
}
\label{fig:angmom}
\end{figure}

Similarly, in Figure~\ref{fig:angmom}, we compare the median angular momentum
of the ejected gas immediately prior to leaving the disk (open
squares), to the median angular momentum of it immediately after reaccretion
(green triangles).  We also denote  the subset of ejected gas that is later re-accreted by the filled red squares.  
There is little distinction between the angular momentum distribution of all the ejected material and the subset of that ejected material that is later reaccreted.  
Hence material that recycles does not have a preferential initial angular momentum, which implies that it does not come from a preferred location in the disk.  
In contrast, it is clear that the reaccreted gas has significantly higher angular momentum at the time of reaccretion than it did when ejected.

In the bottom panel of Figure~\ref{fig:angmom} we plot the ratio of angular momentum of the gas when reaccreted to that when ejected.
This figure shows that ejected gas is re-accreted with higher angular
momentum, typically increased by a factor of $\sim \times 2-3$.
There is a large scatter and no clear trend with mass, although the
highest mass galaxies all have fairly low ``boosting" of the angular
momentum; a larger sample will be necessary to see if that is
statistically significant.  These results are consistent with
previous work by \citet{Brook11b} showing that ejected gas is ``spun
up" in the halo before re-accretion.  It is qualitatively consistent
with the redistribution of angular momentum necessary to match
observations of dwarf galaxy angular momenta by \citet{VanDenBosch01a}.

\section{Discussion}\label{sec:discuss}

The blastwave feedback model used here is essentially a variation on the ``energy-driven" wind scenario. 
In the energy-driven wind model, gas is driven from galaxies assuming SN energy is conserved.  
This scenario is in contrast to the ``momentum-driven" wind model \citep{Murray2005} in which energy may be dissipated away but momentum is conserved.
The difference between the blastwave feedback model and the analytic models for energy driven winds applied to many cosmological simulations is that in the blastwave model the transfer of supernova energy to the interstellar media is based on the local gas properties and is ignorant of the larger galaxy potential.
Despite this key difference, the scaling for mass loading factor as a function of $v_{\rm circ}$ is very close to the analytically derived scaling for energy-driven winds: $v_{\rm circ}^{-2.2}$ versus $v_{\rm circ}^{-2}$.  
We also found that the outflow velocity scalings determined for the sample of simulated galaxies follow the analytic solution for the energy-driven model.
The median wind velocities of the ejected and expelled material had an approximately linear scaling with circular velocity.
This is consistent with the analytic solution for energy-driven winds, namely $v_w = 2\sigma\sqrt{f_L- 1}$~\citep{Murray2005} where $\sigma$ is the velocity dispersion and, therefore, proportional to the circular velocity.

The correspondence between analytic models and the blastwave feedback results imply that the relationship between mass loading and galaxy mass comes directly from the halo potential.
Nevertheless, local gas properties {\em do} affect the efficiency of the energy transfer and, therefore, the scaling of the $v_{\rm circ}$--mass loading relationship.
\citet{Christensen12a} showed that the blastwave feedback model results in a much more efficient removal of gas when combined with a highly-resolved interstellar media that included the cold molecular phase, as used here.
This change in the efficiency is also apparent by comparing the results here to those in \citet{Woods2014}, which examined a blastwave feedback model using lower resolution simulations and without a molecular hydrogen model.
In \citet{Woods2014}, the blastwave feedback model was unable to remove baryons from $\approx 10^{12} \Msun$ halos whereas here 10 -- 20\% of the gas accreted to halos in this mass range were removed by a redshift of zero (see Figure~\ref{fig:baryfrac}).

A somewhat remarkable aspect of the models shown here is that they are able
to match the mass-metallicity relation despite having energy-driven
wind scalings.  In general, cosmological simulations that include
energy-driven winds tend to produce a mass-metallicity relation
that is too steep~\citep{Dave2013,Somerville2015}.  
This may be because such simulations tend to rely solely on ejective feedback to suppress star formation at low masses (above the filtering mass).  
The simulations presented here, though, show that there is
significant preventive feedback, even above the filtering mass, along with a mass-dependent star formation efficiency.
For example, the fraction of halo gas accreted onto the disk is largely independent of halo mass over this sample of galaxies (see Figure~\ref{fig:baryfrac}), indicating that 
the suppression of halo gas accretion onto the disk is similar between dwarf and L$^*$ galaxies.
This preventative feedback can reduce the star formation
in low-mass galaxies.
It does not, however, have a strong impact on the gas-phase
metallicity because that is determined by a competition between the accretion
of fresh gas and the creation of new metals from stars~\citep[e.g.][]{Finlator08}.  Hence it
appears that to simultaneously match the stellar mass function (or
almost equivalently, the stellar mass-halo mass relation) and the
mass-metallicity relation requires having significant preventive
feedback across all mass scales, not just at high masses when active galactic nuclei
feedback putatively happens.  In our simulations, this may happen
owing to wind energy that retards inflow into the disk, as also
seen by \citet{vdv11}.

The energy-driven scaling determined in this suite of galaxy
simulations follows the general convergence within the community
towards energy-driven scalings at lower halo masses \citep{Ford13b,Vogelsberger2014}. 
It should be noted that our model does not include a prescription for different forms of
early stellar feedback, such as radiation pressure, which are generally more consistent with a momentum-driven wind
scenario.  

Different models of early stellar feedback have been incorporated into a variety of galaxy formation codes \citep{Ceverino2013,Aumer2013,Stinson13,Hopkins2013},
and they have been remarkable for their ability to reduce early star formation and to produce galaxies that match the high-redshift stellar mass-halo
mass relation \citep{Agertz2014,Wang2015}.
While very similar versions of the code used here have been shown to produce dwarf galaxies with star formation histories consistent with those observed from resolved stellar populations of local dwarf galaxies \citep{Shen2013}, the larger galaxies likely have too much early star formation \citep{Christensen14}.  
This potential discrepancy suggests that a combination of energy-driven and momentum-driven wind may yet be necessary to increase outflows sufficiently in higher density gas to produce realistic star formation histories.
An alternative to the addition of early stellar feedback exists, however, in improved the modeling of supernova feedback from clusters of young stars.
\citet{Keller2014} describes such a parameter-free, resolution independent supernova feedback model.
Similarly to early stellar feedback models, this model is able to reduce star formation before $z=2$ and to produce a Milky Way-mass galaxy with low bulge-to-total mass \citep{Keller2015}.

In addition to affecting the removal of gas, the numerics of the simulation may affect the rates of reaccretion of gas.  
The blastwave feedback model transfers supernova energy directly to the surrounding gas particles in the
form of thermal energy while simultaneously disabling cooling. As
a result, gas particles may reach very high ($> 10^5$K) temperatures
while temporarily retaining their high density.  The subsequent
reaccretion time scale will be affected by their ability to cool
at high temperatures.  Other feedback models that include less
thermal heating of the gas, such as those using a
hydrodynamically-decoupled kinetic wind model~\citep{SpringelANDHernquist03a},
should be expected to cool and reaccrete more readily.  

Another important numerical factor to consider with examining
reaccretion rates in simulations is the presence of thermal instabilities in the
halo.  As SPH effectively refines in resolution based on density,
the low density regions of the halo are resolved to a lesser degree.
As a result, instabilities and small scale structure will be less
accurately modeled.  Such instabilities become most
sensitive to numerics when a hot gaseous halo is present~\citep{agertz07}.
However, at the masses considered in this study, most simulations suggest
that a virial shock is not able to be supported~\citep[though see
\citealt{Nelson2013}]{Birnboim2003,Keres05}.

The version of SPH used in this paper is known to suffer from
spurious ``surface tension" forces, which make it poor at modeling
Kelvin-Helmholtz instabilities.  However, \citet{Hopkins2013} showed
that the changes due to different SPH modeling were insignificant
compared to the feedback models chosen.  Furthermore, the net amount
of gas accretion (as opposed to its thermal state) appears
to be fairly consistent across different hydrodynamic implementations
\citep{Nelson2013}.  Until a complete suite of high resolution
galaxies can be simulated with a modified version of the SPH formalism
(Wadsley et al. 2016 in prep.) or a new hydrodynamics methodology~\citep[e.g.][]{Hopkins2015}, these simulations represent our
most complete understanding of the scaling outflow properties with
mass.

\section{Conclusions}\label{sec:conclude}

Using high-resolution simulations, we have compared the outflow properties of twenty galaxy halos spanning the mass range from 10$^{9.5}-10^{12} \Msun$ using particle tracking to identify and follow outflowing gas.  
We show that the resulting galaxies match observed global galaxy properties, 
indicating that their baryonic content is realistic in terms of stellar mass, metallicity, and kinematics.  

Since these trends are determined in a large part by the rates of gas inflow and outflow, these simulated galaxies present an opportunity to analyze the properties of outflows that plausibly produce realistic galaxies.  
We tracks gas particles to identify pristine gas accretion, recycled accretion, and gas leaving the disk and/or the halo.  
From this tracking we were able to determine the efficiency of outflows, the relative mass of outflowing and recycled gas, and the properties of the gas both prior to outflow and after accretion.
Since the stellar feedback recipe used in these simulations depended only on the local properties of the ISM (not the host galaxy properties), trends in the outflow properties with halo mass must result from the dynamics of the simulation.

Our conclusions are summarized as follows:

\begin{itemize}
\item With decreasing galaxy mass, galaxies are significantly more efficient at generating outflows.
Specifically, mass loading factors show a power-law dependency on circular velocity, with an exponent of $\approx -2$, which is consistent with energy-driven wind models.  
The similarity between the scaling for the analytic solution and that measured for the simulations argues for the mass loading factor being primarily a function of the global halo properties; the greater efficiency of small galaxies at driving outflows is simply a result of their shallower potential well.
Furthermore, there was no redshift evolution in the scaling of the mass loading factors, which is consistent with $\eta$ being primarily a function of halo depth.

\item In $L^*$ galaxies, ejective feedback is able to reduce the baryonic disk mass by 20\%, while in galaxies with halo masses $\lesssim 10^{10} \Msun$ there can be as large as an 80\% mass reduction.  
These fractions are very similar to the stellar-to-disk-mass ratio across a range of galaxy masses, indicating that ejective feedback is comparable in significance to the globally averaged star formation efficiency in setting the stellar mass of galaxies.
Preventative feedback also plays an important role in setting the stellar mass fraction across the entire mass range; galaxies with masses lower than $10^{11} \Msun$ had reduced baryon masses compared to the cosmic baryon fraction and across the entire mass range only about half of the gas accreted onto the halo was later accreted onto the disk. 

\item 
Recycling is shown to be a common feature of galaxy evolution.
Approximately 50\% of gas that becomes dynamically unbound from the disk (ejected) is later reaccreted across all masses of galaxies.
Such recycling occurs primarily on short time scales.  
The median time scales are $\sim 1$~Gyr, with very little dependence on halo mass, and the timescales follow a logarithmic fall-off.

\csznote{
Decreasing halo mass requires less energetic outflows for escape but less energetic gas is also considered part of the outflow.  
A higher fraction of gas heated by supernovae in dwarf galaxies leaves the disk.  
However, a lower fraction of the gas outflowing from the disk in lower mass galaxies leaves the halo, as the lower baryonic fractions of dwarf
galaxies increase the dynamical range of gas capable of becoming
unbound form the disk while remaining bound to the halo.
}

\item  The source of outflowing material roughly follows the spatial distribution of star formation.
As such, gas is preferentially removed from the centers of galaxies.
Gas is subsequently reaccreted with higher angular momenta and farther out in the disk, indicating that it is ``spun up" through interactions with the halo.
This trend is consistent with previous work indicating that feedback-driven outflows can have a significant impact on the angular momentum distribution of disk baryons.

\end{itemize}

These results give a quantitative picture for how preventive feedback, ejective feedback, and star formation efficiency plausibly combine to yield the baryon fractions in galaxies seen today.
In $L^*$ halos, the $\sim 25$\% efficiency of baryonic conversion into stars is primarily driven by the inability of baryons in the halo to accrete onto the disk, presumably owing to the rapid growth of a hot hydrostatic halo~\citep[e.g.][]{Dekel09,Gabor2014}.  
In modest-sized halos down to $\sim 3\times 10^{10}M_\odot$, ejective feedback becomes the dominant modulator of star formation, as large-scale cosmological models often assume.
At smaller (dwarf) masses, all three effects are comparably important: there is significant prevention of accretion onto the halo, there is an increasing ejection rate, and of the material remaining in the disk, less is formed into stars.  

A large number of the outflow properties we examined are mostly independent of halo mass.  
Lower mass galaxies were more efficient at producing outflows, both in terms of their higher mass loading factors and the higher fractions of disk gas that was ejected.
Indeed, the scaling of the mass loading factor with circular velocity calculated from our simulations is consistent with previous observations \citep{Arribas2014,rupke05,Chisholm2014}.
In contrast, the gas that was ejected from the disk tends to have similar $v/v_{circ}$, recycling times, and re-accreted fractions across all halo masses.
These similarities indicate that reducing the galaxy mass lowers the threshold for driving outflows, but that the outflows themselves
are quite similar across galaxy mass when scaled by the relevant galaxy property.  This is somewhat remarkable given the complex driving mechanisms and interplay of outflowing gas with ambient ISM and halo material.

Together, our results strongly argue for galactic outflows being fundamental to setting the mass of baryons and their distribution within galaxies.
Metallicities offer yet another strong constraint on the history of the baryon cycle.
Future work will analyze the source of metals, their eventual location, and the relative rates of pristine versus enriched accretion.
Such metal distribution mechanisms can be directly constrained by observations of metal absorption lines in quasar spectra passing near galaxies, such as that obtained from the COS-Halos program~\citep{Tumlinson2011,Ford2015}.

\section{Acknowledgements}\label{sec:acknowledge}
The authors are grateful for the referee's helpful suggestions.
They would like to thank Rachel Somerville for her insightful feedback.
These simulations were run at  NASA AMES and Texas Supercomputing Center. 
CC acknowledges support from NSF grants AST-0908499 and AST-1009452.
FG acknowledges support from grant HST GO-1125, NSF AST-0908499, and AST-1410012, and NASA ATP13-0020
TQ acknowledges support from NSF grant  AST-0908499 and NSF AST-1311956.
RD acknowledges support from the South African Research Chairs Initiative and the South African National Research Foundation.
This research was also supported in part by the NSF under Grant No. NSF PHY11-25915
and AST-0847667 and by NASA grant NNX12AH86G

\bibliographystyle{apj} \bibliography{./outflows}

\csznote{
\clearpage
\begin{landscape}
\begin{deluxetable}
\end{deluxetable}
\clearpage
\end{landscape}
}






\end{document}